\documentclass[preprint,12pt]{elsarticle}

\usepackage{amssymb}

\usepackage{graphicx}
\usepackage{newtxtext}
\usepackage{newtxmath}
\usepackage{natbib}
\usepackage{hyperref}
\hypersetup{
	colorlinks = true,
	urlcolor   = blue,
	citecolor  = black,
}
\usepackage[dvipsnames]{xcolor}

\newcommand{\RomanNumeralCaps}[1]
\linenumbers
\usepackage{multirow}
\usepackage{tabularx}

\journal{XXXXX}

\begin{document}

\begin{frontmatter}



\title{Fluid-driven granular dynamics through a consistent multi-resolution particle method}


\author[inst1]{Mojtaba Jandaghian}
\ead{mojtaba.jandaghian@polymtl.ca}
\author[inst1,inst2]{Ahmad Shakibaeinia}
\ead{ahmad.shakibaeinia@polymtl.ca}


\affiliation[inst1]{organization={Department of civil, geological, and mining engineering},
            addressline={Polytechnique Montreal}, 
            city={Montreal},
            country={Canada}}
            
\affiliation[inst2]{organization={Canada research chair in Computational Hydrosystems, Canada}}  

\begin{abstract}
Granular dynamics driven by fluid flow is ubiquitous in many industrial and natural processes, such as fluvial and coastal sediment transport. Yet, their complex multiphysics nature challenges the accuracy and efficiency of numerical models. Here, we study the dynamics of rapid fluid-driven granular erosion through a mesh-free particle method based on the enhanced weakly-compressible Moving Particle Semi-implicit (MPS) method. To that end, we develop and validate a new multi-resolution multiphase MPS formulation for the consistent and conservative form of the governing equations, including particle stabilization techniques. First, we discuss the numerical accuracy and convergence of the proposed approximation operators through two numerical benchmark cases: the multi-viscosity Poiseuille flow and the multi-density hydrostatic pressure. Then, coupling the developed model with a generalized rheology equation, we investigate the water dam-break waves over movable beds. The particle convergence study confirms that the proposed multi-resolution formulation predicts the analytical solutions with acceptable accuracy and order of convergence. Validating the multiphase granular flow reveals that the mechanical behavior of this fluid-driven problem is highly sensitive to the water-sediment density ratio; the bed with lighter grains experiences extreme erosion and interface deformations. For the bed with a heavier material but different geometrical setups, the surge speed and the transport layer thickness remain almost identical (away from the gate). Furthermore, while the multi-resolution model accurately estimates the global sediment dynamics, the single-resolution model underestimates the flow evolution. Overall, the qualitative and quantitative analysis of results emphasizes the importance of multi-scale multi-density interactions in fluid-driven modeling. 
\end{abstract}




\end{frontmatter}


\section{Introduction}\label{sec:intro}
\begin{figure}
	\centering{\includegraphics[width=\textwidth]{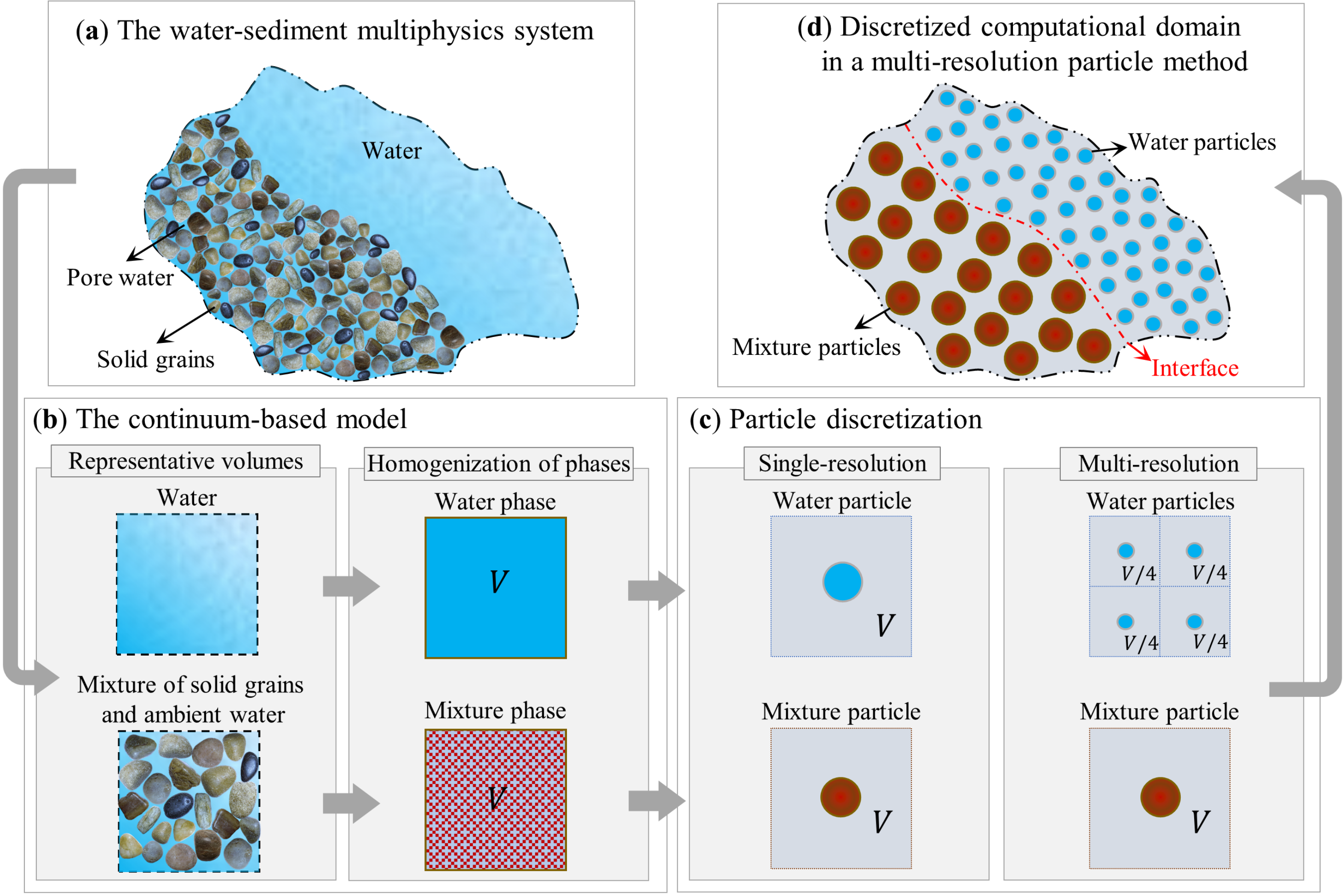}} 
	\caption{ (\textbf{a}): The multiphysics system containing water and submerged solid grains. (\textbf{b}): The  representative volumes of water and mixture phases homogenized into the representative elementary volume, $ V $, in the single-continuum model. (\textbf{c}): The particle discretization in the single- or multi-resolution particle method. (\textbf{d}): The discretized computational domain with different spatial resolutions for the water and mixture particles.}
	\label{fig:1}
\end{figure}

Many hydro-environmental and geotechnical problems involve the multiphase flow of granular material, like sediment, immersed in a fluid, like water (Figure \ref{fig:1}-a). In such multiphysics systems, the granular phase demonstrates dynamic solid- and fluid-like behaviors induced by gravity and the ambient fluid flow (e.g., in the cases of submarine landslides \cite{Robbe2021, Pilvar2019, Bougouin2018, Rondon2011, Balmforth2005} and fluvial and coastal sediment transport \cite{Lobkovsky2008,Fraccarollo2002, Brooks1999}). Particularly, rapid shearing flows over granular beds cause large interfacial deformations leading to the erosion, suspension, and deposition of grains \cite{Guazzelli2018}. The complex and simultaneous presence of the quasi-static, dense-flow, and kinetic (suspended) regimes \cite{Jop2015} makes the accurate prediction of such immersed granular flows challenging. This paper studies the fluid-driven granular dynamics through a consistent numerical method.

Discrete-based and continuum-based methods have been widely developed for simulating immersed granular flows. While discrete-based methods, such as the Discrete Element Method (DEM) \cite{Cundall1979}, provide an in-depth grain-scale understanding of the granular behavior, they are computationally expensive for practical problems that involve a large volume of materials. On the other hand, the continuum-based numerical methods homogenize the assembly of solid grains (or the mixture of solid granules and interstitial fluid) into a body of continuum at the macroscopic level (figure \ref{fig:1}-b). Thus, they are scalable and computationally affordable for large-scale modeling. Such methods employ a rheology model to estimate the mechanical behavior of the granular continuum \cite{Guazzelli2018}.

Mesh-based continuum methods have been developed for various granular simulations \cite{Rauter2021, Lacaze2021}. However, due to their mesh dependency, they require particular treatments to deal with highly dynamic interfaces \cite{Selcuk2021,Rycroft2020}. In contrast, mesh-free Lagrangian continuum methods, or simply particle methods, discretize the continuum using moving particles without any connectivity (Figure \ref{fig:1}-c). This feature of particle methods introduces them as reliable numerical approaches for handling interfacial deformations, and hence, suitable for highly dynamic granular flows \cite{Luo2021, Feng2021, Shakibaeinia_awr2_2012}. The Material Point Method (MPM) \cite{Sulsky1994}, the Moving Particle Semi-implicit (MPS) method \cite{Koshizuka1996}, and the Smoothed Particles Hydrodynamics (SPH) method \cite{Gingold&Monaghan1977} are some of the most widely adopted continuum particle methods. We establish the numerical method of this study based on the MPS formulation. 

SPH and MPS have gone through significant developments to improve their accuracy and stability for highly dynamic and multiphase flows \cite{Antuono2021, Jandaghian2022, Jandaghian2021_3, Rezavand2020, Duan2017, Khayyer2017, Shakibaeinia_cmame_2012}. Shakibaeinia and Jin \cite{Shakibaeinia_wcmps_2010} introduced the weakly compressible MPS method (WC-MPS). Jandaghian and Shakibaeinia \cite{Jandaghian2020} and Jandaghian et al. \cite{Jandaghian2021_3} enhanced the accuracy and stability of the WC-MPS method by proposing artificial diffusive terms and particle regularization techniques. SPH and MPS, coupled with various rheological equations, have simulated immersed granular flows; the adopted constitutive laws include the Bingham plastic formulation \cite{Nabian2017,Khanpour2016, Rodriguez2004}, the Herschel-Bulkley model \cite{Shakibaeinia_awr1_2011, Rodriguez2004}, the Herschel-Bulkley-Papanastasiou model \cite{Fourtakas2016}, the regularized $ \mu(I) $ equation by Job et al. \cite{Jop2006} \cite{Qi2022, Jandaghian_IAHR2019, Tajnesaie2018,Nodoushan2018}, and the elastic-viscoplastic model \cite{Alex2018}. Moreover, to improve the prediction of the incipient motion of the granular particles, some methods employed an additional yielding threshold based on Shields' erosion criterion  \cite{Zubeldia2018,Khanpour2016,Manenti2012}. Recently, \cite{Jandaghian2021_2} proposed a generalized constitutive law to simulate three-dimensional immersed granular collapses and slides. They introduced the regularized form of the visco-inertial rheology model \cite{Baumgarten&Kamrin2019} and the consistent effective pressure term for rapid granular deformation considering non-hydrostatic pore-water pressure and without shear stress threshold. While the previous advanced particle methods have mostly focused on the gravity-driven granular flows, a few have attempted to simulate the fluid-driven cases. 

The continuum-based models treat the dense multiphase granular flow system using either single-continuum or two-continuum models \cite{Guazzelli2018, Jop2015}. The single-continuum models consider the solid grains and interstitial (pore) fluid as a single uniform dense mixture. The mixture interacts with the ambient fluid phase directly through one set of governing equations. The two-continuum models simulate the relative motion of the interstitial fluid and the solid grains and solve separate sets of governing equations (including inter-phase forces) \cite{Guazzelli2018,Pailha2009}. In both approaches, the numerical element representing the continuum (i.e., the representative elementary volume, $ V $, in Figure \ref{fig:1}-b) must be large enough to contain a sufficient number of solid grains. Otherwise, the continuum assumption and the constitutive law would become invalid \cite{Jandaghian2021_2,Guazzelli2018,Rycroft2009}. On the other hand, the numerical element should be small enough (with respect to the characteristic length-scale of the problem) to mathematically represent its vicinity and minimize the numerical approximation errors \cite{Jandaghian2021_2, Alex2018, Baveye1984}. The computational model shall respect these two conflicting constraints in determining the spatial resolution of the discretized domain. Moreover, to capture highly dynamic and high Reynolds number fluid flows, the numerical solution requires a higher spatial resolution for the fluid phase (e.g., $ V/4 $) than that of the granular (or mixture). Accordingly, one can conceive a multi-resolution approach for simulating the fluid-driven granular flows (Figure \ref{fig:1}-c \& d).

Several multi-resolution SPH and MPS methods have been developed for improving numerical accuracy and capturing more accurate flow/solid deformations over a refined computational domain. They implement dynamic particle splitting and merging \cite{Liu2021, Tanaka2019, Vacondio2013}, adaptive particle refinement \cite{Yang2021, Chiron2018, Barcarolo2014}, overlapping methods \cite{Yamada2021, Shibata2017}, and volume adaptive scheme \cite{Sun2020_VAS}. Such multi-resolution methods are employed for various single-phase and multiphase flows (e.g., \cite{Liu2021}, \cite{Sun2018_MR-dplus}, and \cite{Omidvar2013}) and fluid-structure interactions \cite{Khayyer2021_FSI,Khayyer2019_FSI,Zhang2020_FSI,Sun2019_FSI}. Nevertheless, no multi-resolution particle method has been reported for multiphase granular flows.

In the present study, we propose a consistent multi-resolution particle method, based on the enhanced WC-MPS formulation (by Jandaghian and Shakibaeinia\cite{Jandaghian2020} and Jandaghian et al. \cite{Jandaghian2021_2, Jandaghian2021_3}), to investigate the mechanical behavior of immersed granular dynamics primarily driven by rapid fluid flow. We derive new approximation operators respecting the conservative properties of the multi-scale multiphase system. We present the novel multi-resolution MPS method in section \ref{sec:MM_MPS}. To ensure numerical stability and avoid unphysical oscillations (which can largely affect the granular yield behavior), we adapt the modified diffusive term \cite{Jandaghian2021_2} and the Dynamic Particle Collision (DPC) technique \cite{Jandaghian2021_3} to the newly developed multi-resolution framework. Further, we couple the developed MPS method with the generalized rheology model by Jandaghian et al. \cite{Jandaghian2021_2} supplied with a suspension equation for rapid and immersed granular flows (section \ref{sec:GRM}). First, we validate the numerical model to two multiphase benchmark cases, i.e., the multi-viscosity Poiseuille flow and the multi-density hydrostatic pressure, quantifying the numerical accuracy and convergence of the multi-resolution formulation (section \ref{sec:Converg}). Next, by simulating water dam-break waves on erodible sediment beds, we provide a comprehensive study on the dynamics of rapid sediment erosion induced by a sudden collapse of a water column (section \ref{sec:EDB}). Validating the numerical solutions with the available experimental data \cite{Spinewine2007, Spinewine2013}, we present internal flow properties and the global mechanical behavior of this fluid-driven problem. We also parameterize the rheology and analyze the phenomenology of sediment dynamics concerning different bed materials and initial configurations of the test case. Comparing the results of the single- and multi-resolution MPS simulations, we evaluate the role of multi-scale interactions in capturing the flow evolution.  

\section{Equations of motion}\label{sec:GovEq}
Here, the system of granular material and the ambient fluid is considered as a multiphase continuum, which in a Lagrangian framework is described by the continuity equation:
\begin{equation} \label{eq:Cont}
	\frac{\mathrm{D}\rho}{\mathrm{D}{t}}=-{\rho}\nabla\cdot\boldsymbol{v},
\end{equation}
the momentum equation:
\begin{equation} \label{eq:Mom}
	\frac{\mathrm{D}\boldsymbol{v}}{\mathrm{D}{t}}=\frac{\nabla\cdot{T}}{\rho}+{\boldsymbol{F}},
\end{equation}
and the advection equation:
\begin{equation} \label{eq:Adv}
	\frac{\mathrm{D}\boldsymbol{r}}{\mathrm{D}{t}}={\boldsymbol{v}},
\end{equation}
calculating the time evolution (${{D}(.)}/{{D}{t}}$) of the fluid density, ${\rho}$, velocity, ${\boldsymbol{v}}$, and position, ${\boldsymbol{r}}$, respectively \cite{Panton2013}. The total stress tensor, $ {T} $, consists of the pressure scalar value, ${p}$, and the shear stress tensor, ${t}$, as ${T}=-{p}{{I}}+{{t}}$ (${{I}}$ being the identity matrix) and ${\rho}{\boldsymbol{F}}$ is the body force per unit volume. Considering the barotropic fluids, the equation of state calculates the pressure, i.e., ${p}={f}({\rho})$. For incompressible fluid flows and by neglecting the gradient of dynamic viscosity, ${\eta}$, over the fluid domain (i.e., considering $\nabla\cdot\boldsymbol{v}\approx 0$ and $\nabla{\eta}\approx0$), the divergence of the stress tensor reduces to:
\begin{equation}\label{eq:Shear}
	{\nabla\cdot{T}}=-{\nabla{p}}+{\eta}{\nabla^2\boldsymbol{v}}
\end{equation}
which is valid for Newtonian and non-Newtonian mechanical behaviors. Constitutive laws determine ${\eta}$ as a function of hydrodynamic and material characteristics \cite{Guazzelli2018}. For water, we include a simple turbulence model (based on the Large Eddy Simulations) in the shear force calculations. We treat the mixture phase as a non-Newtonian fluid through the visco-inertial rheology model (proposed by Baumgarten and Kamrin \cite{Baumgarten&Kamrin2019}, and then represented in the regularized form by Jandaghian et al. \cite{Jandaghian2021_2}).

\section{A consistent multi-resolution multiphase MPS method}\label{sec:MM_MPS}
\subsection{Integral and summation interpolants}\label{sec:IntSum}
In continuum-based particle methods, the approximation operator transforms the integral representation of functions into the summation interpolant. By discretizing the computational domain, $ {\Omega} $, moving calculation points (or simply particles) carry flow and material properties \cite{Liu2003}. Here, we adopt the general integral formulation of the MPS method to derive the summation operator of the multi-resolution model. 

MPS integral representation of an arbitrary function, $ {f(\boldsymbol{r})} $, reads \cite{Koshizuka1998}:
\begin{equation}\label{eq:intfW}
	{f(\boldsymbol{r})}=\frac{\int_{\Omega}f(\boldsymbol{r}^{\prime})W(\lVert\boldsymbol{r}^{\prime}-\boldsymbol{r}\rVert, r_{e})dr^{\prime}}{\int_{\Omega}W(\lVert\boldsymbol{r}^{\prime}-\boldsymbol{r}\rVert, r_{e})dr^{\prime}}
\end{equation}
where $ {dr^{\prime}} $ is a differential volume element. The positive non-dimensional weighting function, $ {W} $, (so-called the kernel) with a compact support smooths $ {f} $ over the influence radius, $ {r}_{e} $. MPS introduces a normalization factor, denoted by $ {n_{0}} $, into the equations which corresponds to the reference physical fluid density, $ {\rho_{0}} $, and the mass of volume element, $ {m} $, via:
\begin{equation}\label{eq:n0}
	{n_{0}}=\frac{{\rho}_{0}}{m}\int_{\Omega}W(\lVert\boldsymbol{r}^{\prime}-\boldsymbol{r}\rVert, r_{e})dr^{'}.
\end{equation}
Nevertheless, $ n_0 $ only depends on the type of kernel and the ratio of $ {r}_{e} $ to the size of spatial discretization, $ {l_0} $ (i.e., $ {k}={{r}_{e}}/{l_0} $) \cite{SoutoIglesias2013}. With considering a constant $ {k} $ for the multi-resolution model, we employ (\ref{eq:n0}) to rewrite (\ref{eq:intfW}) as:
\begin{equation}\label{eq:intfW2}
	{f(\boldsymbol{r})}=\frac{1}{n_{0}}{\int_{\Omega}{f}(\boldsymbol{r}^{\prime})\frac{W(\lVert\boldsymbol{r}^{\prime}-\boldsymbol{r}\rVert, r_{e}){k^{d}}}{{{r}_{e}}^{d}}dr^{\prime}}
\end{equation}
in which $ {d} $ is the number of space dimensions and $ {m}/{{\rho}_{0}}=({r_{e}}/{k})^{d} $. 

The original MPS formulation derives the summation operator of the integral representation (\ref{eq:intfW2}) by considering identical smoothing length and spatial resolution for the entire fluid domain. Here, we propose a new formulation for the kernel, $ \widetilde{W} $, to account for multi-resolution particle interactions through the general form of approximation operator:
\begin{equation}\label{eq:SumfW}
	{\langle{f}\rangle}_{i}=\frac{1}{n_{0}}{\sum_{i\neq j}^{N} {f}_{j}\widetilde{W}({r}_\mathit{ij},\overline{r_{e}}_\mathit{ij}, {V_0}_{j})}
\end{equation} 
for a target particle, $ {i \in \Omega} $, surrounded by $ {N} $ number of neighbour particles, identified as $ {j\in \Omega} $, where $ {r}_\mathit{ij}={\lVert\boldsymbol{r}_{j}-\boldsymbol{r}_{i}}\rVert\leq {r_{e}}_{i}$. The modified kernel, which is non-dimensional, would be:
\begin{equation}\label{eq:What}
	\widetilde{W}_\mathit{ij}=\widetilde{W}({r}_\mathit{ij},\overline{r_{e}}_\mathit{ij}, {V_0}_{j})=\frac{W({r}_\mathit{ij}, \overline{r_{e}}_\mathit{ij}){V_0}_{j}{k^{d}}}{\overline{r_{e}}^{d}_\mathit{ij}},
\end{equation} 
in which $ {V_0}_{i} $, being the reference volume of particle, is equal to $ {({l_0}_{i})^{d}} $ for incompressible fluid flows (as $ {l_0}_{i} $ stands for the initial particle spacing of $ {i} $). To respect the symmetric feature of the smoothing procedure in the governing equations (see section \ref{sec:Disc} and Figure \ref{fig:Kern}), we have substituted $ {r_{e}} $ with $ {\overline{r_{e}}_\mathit{ij}}={({r_{e}}_{i}+{r_{e}}_{j})/2} $ (similar to the formulations used in \cite{Sun2017_dplus} and \cite{Tanaka2019}). The new definition of kernel (\ref{eq:What}) includes the various sizes and smoothing lengths of particles within the approximation operator required for considering the multi-resolution interactions (where $ {V_0}_{i}\neq {V_0}_{j}$); while in the same resolution interactions (where $ {V_0}_{i}={V_0}_{j}$), (\ref{eq:What}) automatically reduces to its original shape as $ \widetilde{W}_\mathit{ij}=W({r}_\mathit{ij}, {r_{e}}_{i}) $. 
By neglecting the kernel truncations at the interfaces and away from boundaries, $ {n_{0}} $ keeps its standard definition as the summation of kernel at the initial uniform distribution of particles, i.e., $ {n_{0}}= \max{\sum_{i\neq j}^{N} {W}({r}_\mathit{ij}, {r_{e}}_{i})} $ at $ t=0 $. Thus, it can be identified as a global constant for all the particle sizes and their interactions (as $ {k={r_e}_i/{l_0}_i} $ is invariable over $ \Omega $). In this study, we set $ k=3.1 $ and use the third-order polynomial spiky kernel function \cite{Shakibaeinia_wcmps_2010} for all the approximation operators.
\subsection{The discrete system of flow equations}\label{sec:Disc}
In particle methods, the moving particles are the representative elementary volume of the ambient water phase, $ \Omega_w $, or the mixture of pore water and solid grains, $ \Omega_m $, or the solid walls,  $ \Omega_s $, forming the computational domain, $ \Omega $ (where $ \Omega_w\cup\Omega_m $ would be the fluid phase, $ \Omega_f$, and $ \Omega=\Omega_f\cup\Omega_s $). Using the summation operator for a target particle $ i\in \Omega_f $,  the flow equations read:  

\begin{equation} 
	\left\{
	\begin{array}{l}
		\displaystyle
		\frac{1}{n_i}\frac{\mathrm{D}{n_i}}{\mathrm{D}{t}}=-\langle{\nabla\cdot\boldsymbol{v}}\rangle_i+{D^m_i}\\[10pt]
		\displaystyle
		{}\frac{\mathrm{D}\boldsymbol{v}_i}{\mathrm{D}{t}}=-\frac{1}{{\rho}_i}\langle\nabla{p}\rangle_i+\frac{1}{{\rho}_i}\langle{\eta}{\nabla^2\boldsymbol{v}}\rangle_i+{\boldsymbol{F}_i}\\[10pt]
				\displaystyle
		\frac{\mathrm{D}\boldsymbol{r}_i}{\mathrm{D}{t}}={\boldsymbol{v}_i},
	\end{array}
	\right.\label{eq:AppGov}
\end{equation}

in which $ {n_{i}} $ is the non-dimensional particle number density (given as $ {n_{0}\rho_{i}/{\rho_0}_i} $) and independent of the density discontinuity at the interfaces \cite{Jandaghian2021_2}. In this model, the momentum equation considers the density of particle, $ \rho_i $, to be equal to the reference density of the fluid phase respecting the original form of the incompressible MPS method (i.e., in the momentum equation: $ \rho_i={\rho_0}_i$ where for $i\in \Omega_w \rightarrow {\rho_0}_i={\rho_0}_w $ and $i\in \Omega_m \rightarrow {\rho_0}_i={\rho_0}_m= {\rho_0}_w(1-\phi_0)+\phi_0 \rho_g $ as $ \phi_0 $ and $ \rho_g $ are the reference volume fraction and the true density of the solid grains, respectively).

For the multi-resolution multi-phase MPS model, we use the kernel (\ref{eq:What}) to discretize the right-hand side terms of (\ref{eq:AppGov}) based on the conservative WC-MPS formulation \cite{Jandaghian2020}:
\begin{equation} 
	\left\{
	\begin{array}{l}
		\displaystyle
		\langle{\nabla\cdot\boldsymbol{v}}\rangle_i=\frac{d}{n_{0}}{\sum_{i\neq j}^{N} \left(\frac{n_{j}}{n_{i}}\right) \frac{\boldsymbol{v}_{j}-\boldsymbol{v}_{i}}{r_\mathit{ij}}\cdot\boldsymbol{e}_\mathit{ij} \widetilde{W}_\mathit{ij}}\\[10pt]
		\displaystyle
		{\langle\nabla{p}\rangle_i}=\frac{d}{n_{0}}{\sum_{i\neq j}^{N} \left({n_{i}}\frac{{p}_{j}}{n_{j}}+{n_{j}}\frac{{p}_{i}}{n_{i}}\right) \frac{\boldsymbol{e}_\mathit{ij}}{r_\mathit{ij}}\widetilde{W}_\mathit{ij}}\\[10pt]
		\displaystyle
		\langle{\eta}{\nabla^2\boldsymbol{v}}\rangle_i=\frac{2d}{n_{0}}{\sum_{i\neq j}^{N} {\eta}_\mathit{ij} \frac{\boldsymbol{v}_{j}-\boldsymbol{v}_{i}}{r_\mathit{ij}^{2} }\widetilde{W}_\mathit{ij}}.
	\end{array}
	\right.\label{eq:AppTerms}
\end{equation}

\begin{figure}
	\centering{\includegraphics[width=5 in]{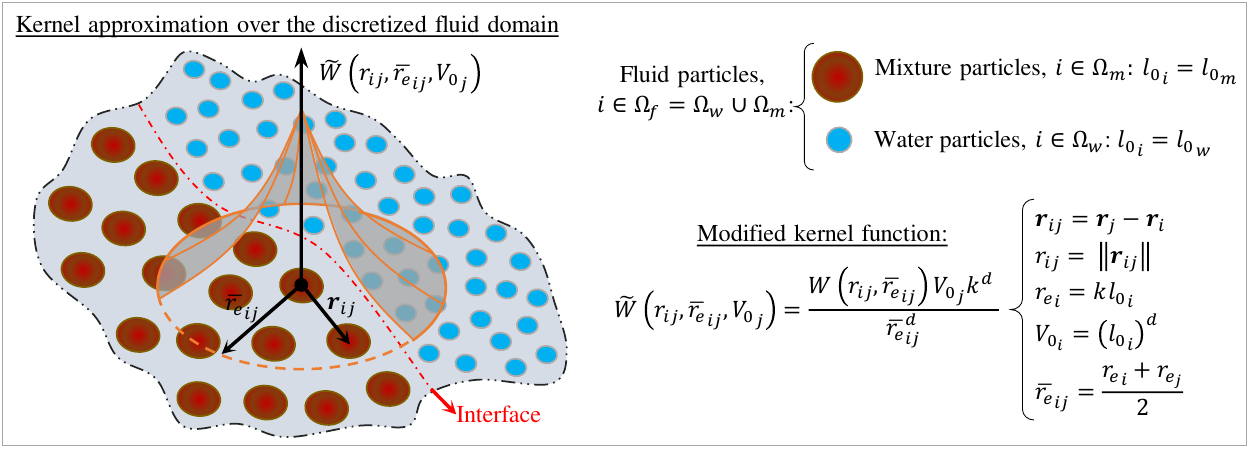}} %
	\caption{Kernel function of the MPS method for the multi-resolution particle interactions. $ d $ is the space dimension equal to 2 and 3 for the two- and three-dimensional simulations, respectively. $ k $ determines the smoothing length of the kernel, $ {r_e}_i $, and is set to 3.1 regardless of the particle size, $ {l_0}_i$.}
	\label{fig:Kern}
\end{figure}

$ {\boldsymbol{e}_\mathit{ij}}={\boldsymbol{r}_\mathit{ij}}/{r_\mathit{ij}} $ is the unit direction vector and the harmonic mean of the dynamic effective viscosity of $ {i} $ and $ {j} $ (i.e., $ \eta_{i}$ and $\eta_{j} $) gives $ \eta_\mathit{ij}=2\eta_{i}\eta_{j}/(\eta_{i}+\eta_{j}) $. With the new kernel function the interaction of particles with various size and density remains anti-symmetric within the governing equations; thus, the conjugate gradient and divergence operators ensure the conservation of the total energy (in the absence of shear and external forces) (see \cite{Price2012} and \cite{Jandaghian2020}) (Figure \ref{fig:Kern}).

Considering the barotropic fluid as a weakly compressible phase, we employ the equation of state to calculate the pressure by \cite{Shakibaeinia_wcmps_2010}:
\begin{equation}\label{eq:EOS}
	p_{i}=B_0 \left(\left(\frac{n_i}{n_0}\right)^\gamma-1\right)
\end{equation}
where the bulk modulus, $ B_0={c_0}^2 {\rho_0}/\gamma $ and $ \gamma=7 $ are constant for all the fluid phases as $ \rho_0 $ and $ c_0 $ are the true density and the artificial sound speed of the reference phase,  respectively (here, we consider water as the reference phase, thus, $ \rho_0={\rho_0}_w $ and $ c_0={c_0}_w $). To limit the compressibility to less than $ 1\% $, the reference sound speed should satisfy $ c_0\geq10{\lVert{\boldsymbol{v}\rVert}_{max}} $ condition by which the Mach number is kept less than 0.1 (${\lVert{\boldsymbol{v}\rVert}_{max}} $ being the maximum expected velocity magnitude).

Next, we adapt the modified diffusive term of Jandaghian et al. \cite{Jandaghian2021_3} to the multi-resolution framework with the new kernel (\ref{eq:What}) as follows:
\begin{equation}\label{eq:Diff}
	{D^m_i}=\left({\delta_{MPS} \frac{\Delta{t} {c^2_0}}{n_0}} \right) \frac{2d}{n_{0}}{\sum_{i\neq j}^{N} \left[(n_j-n_i)-\frac{1}{2}[\langle\nabla{n}\rangle^c_i+\langle\nabla{n}\rangle^c_j]\cdot{\boldsymbol{r}_\mathit{ij}}\right]
		\frac{\widetilde{W}_\mathit{ij}}{r^2_\mathit{ij}}}
\end{equation}
in which $ \langle\nabla{n}\rangle^c_i $ is the high-order gradient operator of $ n_i $ estimated by:
\begin{equation}\label{eq:GradPNUM}
	\langle\nabla{n}\rangle^c_i=\frac{d}{n_{0}}{\sum_{i\neq j}^{N} \frac{{n}_{j}-{n}_{i}}{r_\mathit{ij}}({C}_i\boldsymbol{e}_\mathit{ij}) \widetilde{W}_\mathit{ij}},
\end{equation}
and the correction matrix, $ {{C}_i} $, is given as:
\begin{equation}\label{eq:Ci}
	{{C}_i}=\left(\frac{d}{n_{0}}{\sum_{i\neq j}^{N} \frac{\boldsymbol{r}_{j}-\boldsymbol{r}_{i}}{r_\mathit{ij}}\otimes\boldsymbol{e}_\mathit{ij} \widetilde{W}_\mathit{ij}}\right)^{-1}.
\end{equation} 
where $ \otimes $ stands for the outer product of vectors. The non-dimensional coefficient, $ 0 \le \delta_{MPS}\leq1 $, the calculation time step, $ {\Delta t} $, and $ c_0 $ adjust the magnitude of this numerical correction. The diffusive term obeys the mass conservation law, if $ \sum_i {n_i} {V_i} {D^m_i}=0$ (as $ {V}_i={n_0}{V_0}_i/n_i $ from (\ref{eq:n0})); with $ {i}, {j} \in\Omega_f $ in (\ref{eq:Diff}-\ref{eq:Ci}), the diffusive term would be an anti-symmetric formulation which conserves the total mass of the multi-resolution multi-phase system.
\subsection{Dynamic particle collision for multi-scale multiphase interactions}\label{sec:DPC}
Here, we implement the Dynamic pair-wise Particle Collision (DPC) method (proposed by Jandaghian et al. \cite{Jandaghian2021_3}) as the particle regularization technique which ensures the numerical stability by eliminating particle clustering and high-frequency pressure noises. Considering the velocity variation of two particles colliding with different masses and volumes, we develop the DPC formulation for the multi-resolution multi-phase interactions as:
\begin{equation}\label{eq:DPC}
	\delta \boldsymbol{v}_i=\left(\sum_{i\neq j}^{N}\kappa_\mathit{ij}\frac{2{m_0}_\mathit{j}}{{m_0}_\mathit{i}+{m_0}_\mathit{j}}\boldsymbol{v}^{\mathit{coll}}_\mathit{ij}-\frac{\Delta t}{{\rho_0}_i}\sum_{i\neq j}^{N}\alpha_\mathit{ij}\frac{2{V_0}_\mathit{j}}{{V_0}_\mathit{i}+{V_0}_\mathit{j}}\frac{p^b_\mathit{ij}}{r_\mathit{ij}}\boldsymbol{e}_\mathit{ij}\right)
\end{equation} 
where $ i $, $ j\in\Omega_f $ and $ {m_0}_i={\rho_0}_i {V_0}_i $. The collision velocity, $ \boldsymbol{v}^\mathit{coll}_{ij} $, and the binary multiplier, $ \alpha_{ij} $, are given by:
\begin{equation}\label{eq:VColl}
	(\boldsymbol{v}^\mathit{coll}_\mathit{ij}, \alpha_\mathit{ij}) = \left\{
	\begin{array}{ll}
		\left((\boldsymbol{v}_\mathit{ij} \cdot\boldsymbol{e}_\mathit{ij})\boldsymbol{e}_\mathit{ij},0\right), & \text{for } \boldsymbol{v}_\mathit{ij} \cdot\boldsymbol{e}_\mathit{ij}< 0 \\[2pt]
		\left( 0, 1 \right)         & \text{otherwise}
	\end{array} \right.
\end{equation}
and the dynamic background pressure, $ p^b_{ij} $, is defined as $ p^b_\mathit{ij}=\tilde{p}_\mathit{ij}\chi_\mathit{ij}$ where
$\tilde{p}_\mathit{ij}= \\ \max\left( \min\left( \lambda\left\vert p_{i} + p_{j}\right\vert,\lambda p_{\mathit{max}} \right), p_{\mathit{min}}\right)$, and, $\chi_{ij} = \left({W ( r_\mathit{ij}, \overline{l_0}_\mathit{ij})}/{W( 0.5\overline{l_0}_\mathit{ij}, \overline{l_0}_\mathit{ij})}\right)^{0.5}$. The non-dimensional variable, $ \chi_\mathit{ij} $, is a function of the kernel with the smoothing length set to $ \overline{l_0}_\mathit{ij}= ({l_0}_\mathit{i}+{l_0}_\mathit{j})/2 $ (where for $ r_\mathit{ij}\geq\overline{l_0}_\mathit{ij} \rightarrow \chi_\mathit{ij}=0 $). The preset maximum and minimum pressure of the test case ($ p_\mathit{max} $ and $ p_\mathit{min} $ respectively) and the non-dimensional constant, $ \lambda $, specify the strength of the repulsive term. For the collision term, the variable coefficient, $ \kappa_\mathit{ij} $, dynamically sets the coefficient of restitution as a function of $ r_\mathit{ij} $ via:
\begin{equation}\label{eq:kappa}
	\kappa_\mathit{ij}=\left\{
	\begin{array}{ll}
		\chi_\mathit{ij}&0.5 \overline{l_0}_\mathit{ij} \leq r_\mathit{ij} < \overline{l_0}_\mathit{ij}\\[5pt]
		1 &      r_\mathit{ij} <0.5 \overline{l_0}_\mathit{ij}.
	\end{array} \right.
\end{equation}
Eventually, $ \delta \boldsymbol{v}_i $ from (\ref{eq:DPC}) updates the velocity and position of particles within the solution algorithm (i.e., we have $ \boldsymbol{v}^\prime_i=\boldsymbol{v}_i+\delta \boldsymbol{v}_i $ and $ \boldsymbol{r}^\prime_i=\boldsymbol{r}_i+\delta \boldsymbol{v}_i\Delta t $ ). The proposed DPC through equations (\ref{eq:DPC}-\ref{eq:kappa}) conserves the linear momentum of the multi-resolution multi-phase particle interactions (i.e., with $ i,j\in\Omega_f $ then $ \sum_{i}{m_0}_i\delta\boldsymbol{v}_i=0 $). In this study, we use the Wendland kernel for $ \chi_\mathit{ij} $ and set $ \lambda=0.2 $ \cite{Jandaghian2021_3}.
\subsection{Generalized rheology model}\label{sec:GRM}
We employ the generalized rheology model of Jandaghian et al. \cite{Jandaghian2021_2} for calculating the effective viscosity of the water and mixture particles. For water as a Newtonian fluid with the true viscosity, ${\mu_w}$, the effective viscosity increases by the presence of solid grains (i.e., with the approximated volume fraction, $\langle{\phi}\rangle_{i}$) and including the turbulence effect:
\begin{equation}\label{eq:mu_w}
	{i\in \Omega_w} \rightarrow {\eta_i}={\mu_w}\left(1+\frac{5}{2}\langle{\phi}\rangle_{i}\right)+{{\rho_0}_w}{\nu_t}_i.
\end{equation}
as the eddy viscosity is given by $ {\nu_t}_i=({C_s}{r_e}_i)^2 |\dot{\gamma}|_i$ and the Smagorinsky constant coefficient is set to $ C_s=0.12 $. The pressure-imposed rheology treats the mixture of water and solid grains as a non-Newtonian fluid through the mixture effective viscosity formulated by $ {\eta_i}={{\mu}_i{p_g}_i}/{|\dot{\gamma}|_i} $ in which ${p_g}_i$ is the solid grains' normal stress (i.e., the effective pressure), $ {\mu_i} $ is the friction coefficient, and ${|\dot{\gamma}|_i}$ is the magnitude of strain rate tensor \cite{Guazzelli2018}. With the visco-inertial model of Baumgarten and Kamrin \cite{Baumgarten&Kamrin2019} (as the friction coefficient) and the regularized formulation (for avoiding the singularity issue when $ |\dot{\gamma}|_i=0$), Jandaghian et al. \cite{Jandaghian2021_2} represented the effective viscosity of the mixture particles as:
\begin{eqnarray}\label{eq:mu_m}
	{i\in \Omega_m} \rightarrow {\eta_i}=
	\frac{{\tau_y}_i}{\sqrt{|\dot{\gamma}|^2_i+\lambda^2_r}}+
	\frac{({\mu_2}-{\mu_1}){p_g}_i}{{b}\sqrt{{p_g}_i}/\sqrt{d^2_g\rho_g+{2{\mu_w}}/(|\dot{\gamma}|_i+\lambda_r)}+|\dot{\gamma}|_i}\nonumber\\	 +\frac{5\langle{\phi}\rangle_{i}}{2a}\left(\frac{{\mu_w}{\sqrt{{p_g}_i}}}{\sqrt{|\dot{\gamma}|^2_id^2_g\rho_g+2{\mu_w}|\dot{\gamma}|_i+\lambda^2_r}}\right)
\end{eqnarray}
where $ {a} $ and $ {b} $ are material constants. The upper and lower limits of the solid grains' friction are denoted as $ {\mu_2} $ and $ {\mu_1}=\mathrm{tan}(\theta) $, respectively. $ {\rho_g} $, $ {d_g} $, and $ \theta $ stand for the true density, the mean diameter, and the internal friction angle of the solid grains, respectively. The yield stress, $ \tau_y $, is given by the Drucker-Prager yield criteria as $ {\tau_y}_i=2\sqrt3 \mathrm{sin}(\theta) {p_g}_i/ (3-\mathrm{sin}(\theta))$ noting that $ {p_g}_i>0 $. The regularization parameter, $ \lambda_r $, is set to 0.001. For incompressible fluid flows, ${|\dot{\gamma}|_i}=\sqrt{4{II_E}_i}$, as the strain rate tensor, ${{E}_i}=0.5[{\langle\nabla\boldsymbol{v}\rangle^c_i+({\langle\nabla\boldsymbol{v}\rangle^c_i})^{\text{t}}}]$ and its second principal invariant, ${II_E}_i={0.5}{{E}_i}:{{E}_i}$. For the derivation of (\ref{eq:mu_m}) readers are referred to Jandaghian et al. \cite{Jandaghian2021_2}. We estimate the gradient of velocity and the volume fraction of the water and mixture particles through:	
\begin{equation}\label{eq:GradV}
	\langle\nabla\boldsymbol{v}\rangle^c_i=\frac{d}{n_{0}}{\sum_{i\neq j}^{N} \frac{\boldsymbol{v}_{j}-\boldsymbol{v}_{i}}{r_\mathit{ij}}({C}_i\boldsymbol{e}_\mathit{ij}) \widetilde{W}_\mathit{ij}}
\end{equation}
and
\begin{equation}\label{eq:Phi}
	\langle{\phi}\rangle_{i}=\frac{\sum_{j}^{N} \phi_j\widetilde{W}_\mathit{ij}}{\sum_{j}^{N} \widetilde{W}_\mathit{ij}},
\end{equation}
respectively, noting that for $ j\in \Omega_w \rightarrow \phi_j=0$ and $ j\in \Omega_m \rightarrow \phi_j=\phi_0$.

The non-dimensional parameters, i.e., the inertial number, $ I_i={|\dot{\gamma}|_i}{d_g}\sqrt{\rho_g/{p_g}_i} $, the viscous number, $ {I_\nu}_i=\mu_w{|\dot{\gamma}|_i}/{p_g}_i $, and the mixed number, $ I_m=\sqrt{I^2+2I_\nu} $ govern the visco-inertial rheology model \cite{Amarsid2017, Boyer2011}. This model is validated against the experimental data of immersed granular flows where $ I_m\le0.6 $ \cite{Baumgarten&Kamrin2019}. In rapid fluid-driven granular erosion, mixture particles at the interface are subjected to high shear forces leading to their suspension with low volume concentration. In the dilute and semi-dilute conditions, the dynamic viscosity turns to be a function of the volume fraction and independent of the shear rate magnitude \cite{Guazzelli2018}. Thus, to incorporate the role of suspension effects, we calculate the effective viscosity of the mixture particles where  $ \langle{\phi}\rangle_{i}/\phi_0 < 0.5 $ or $ {I_m}_i>0.6 $ through the suspension equation of \cite{Vand1948}:
\begin{equation}\label{eq:sus}
	{\eta_i}=\mu_w \exp\left(\frac{2.5\langle{\phi}\rangle_{i}}{1-\frac{39}{64}\langle{\phi}\rangle_{i}}\right).
\end{equation}
Coupling the visco-inertial formulation with the Vand's equation aims at simulating the suspension process of mixture particles. Previously, Zubeldia et al. \cite{Zubeldia2018} and Fourtakas and Rogers \cite{Fourtakas2016} used this equation with the Herschel-Bulkley-Papanastasiou constitutive model for sediment dynamics modeling in SPH. One should not that our implemented constitutive model treats the different regimes of the immersed granular flow through the failure and post-failure terms and the suspension equation without any shear stress threshold (e.g., Shield's erosion criterion in \cite{Zubeldia2018, Khanpour2016}) to distinguish the yielded particles from the un-yielded ones.

Moreover, we implement the consistent effective pressure, $ p_{\text{eff.}_i} $, proposed by Jandaghian et al. \cite{Jandaghian2021_2} to estimate $ {{p_g}_i} $ of the immersed granular flow where for $ i\in \Omega_m$, $ {p_g}_i=p_{\text{eff.}_i} $ and,
\begin{equation}\label{eq:Peff}
	p_{\text{eff.}_i}=B_0 \left[\left(\frac{n_i}{n_0}\right)^\gamma-\left(\frac{{\rho_w}_i}{{\rho_0}_w}\right)^\gamma\right]
\end{equation}
in which the density of the pore-water, $ {{\rho_w}_i} $, is updated by its continuity equation derived for the single-phase continuum model as follows:
\begin{equation}\label{eq:pw_cont}
	\frac{1}{{\rho_w}_i}\frac{\mathrm{D}{\rho_w}_\mathit{i}}{\mathrm{D}{t}}=-\langle{\nabla\cdot\boldsymbol{v}}\rangle_i+{D^m_i}.
\end{equation}
The right-hand side of (\ref{eq:pw_cont}) is identical to the right-hand side of the continuity equation used for updating $ n_i $ (in (\ref{eq:AppGov})). 

\subsection{Boundary conditions and solution algorithm}\label{sec:BCTI}

In the numerical model, the fixed boundary particles ($i\in \Omega_s $) simulate the solid walls. The fluid particles interact with the solid boundary particles through the governing equations (\ref{eq:AppTerms}) \cite{Jandaghian2021_2}. To update the pressure of the wall boundary particles ($ p_i $), we implement the dynamic solid boundary condition by Crespo et al. \cite{Crespo2015}. The pressure of the closest wall particle is assigned to the ghost particles. The velocity of the solid boundary particles, $ \boldsymbol{v}_i $, is considered to be zero in the continuity equation of fluid particles. In the shear force calculations, the velocity assigned to the solid boundary particles applies slip or no-slip boundary conditions. For viscous flow simulations, we consider the viscosity of the fluid particle for the boundary particle (i.e., $ \eta_{j\in\Omega_s}=\eta_{i\in\Omega_f} $ in $ \langle{\eta}{\nabla^2\boldsymbol{v}}\rangle_i $).  

For solving the governing equations, we implement the second-order and explict symplectic time integration scheme (represented by Jandaghian et al. \cite{Jandaghian2021_2}). The time step of calculation, $ \Delta t $, is given based on the the Courant–Friedrichs–Lewy (CFL) stability condition and the shear force corresponding to the density, the spatial resolution and the dynamic viscosity of each phase (i.e., $ \Omega_w $ and $ \Omega_m $) as follows:
\begin{equation}\label{eq:DT}
	\Delta t=\text{min}\left\{C_\mathit{CFL}\frac{l_0}{c_0}, C_\mathit{v}\frac{{{\rho_0}}{l_0}^2}{\eta_\mathit{max}}\right\}_{\Omega_{w,m}},
\end{equation}
in which $ C_\mathit{CFL} $ and $ C_\mathit{v} $ are non-dimensional coefficients of the time steps (identical for both phases) and $ \eta_\mathit{max} $ is the maximum expected dynamic viscosity. Considering $ \Omega_w $ as the reference phase for the bulk module (in (\ref{eq:EOS})), we set the sound speed of the second phase as $ {c_0}_m={c_0}_w\sqrt{{\rho_0}_w/{\rho_0}_m} $.
\section{Results and discussions}\label{sec:Res}
The reliability of water-sediment dynamics modeling depends on the accuracy of the approximated governing equations and their capability in capturing the multiphysics flow properties. To investigate the consistency of the proposed multi-resolution MPS formulation, we begin with studying the numerical accuracy and convergence of two benchmark cases, i.e., the multi-viscosity Poiseuille flow and the hydrostatic pressure of two fluid phases (section \ref{sec:Converg}). Then, we investigate and validate rapid fluid-driven granular erosion through simulating dam break waves over movable beds (section \ref{sec:EDB}). A movie containing the numerical simulations and results is provided as the supplementary data of this paper. 

\subsection{Numerical accuracy and convergence of the multi-resolution MPS model}\label{sec:Converg}
\begin{figure}
	\centering{\includegraphics[width=\textwidth]{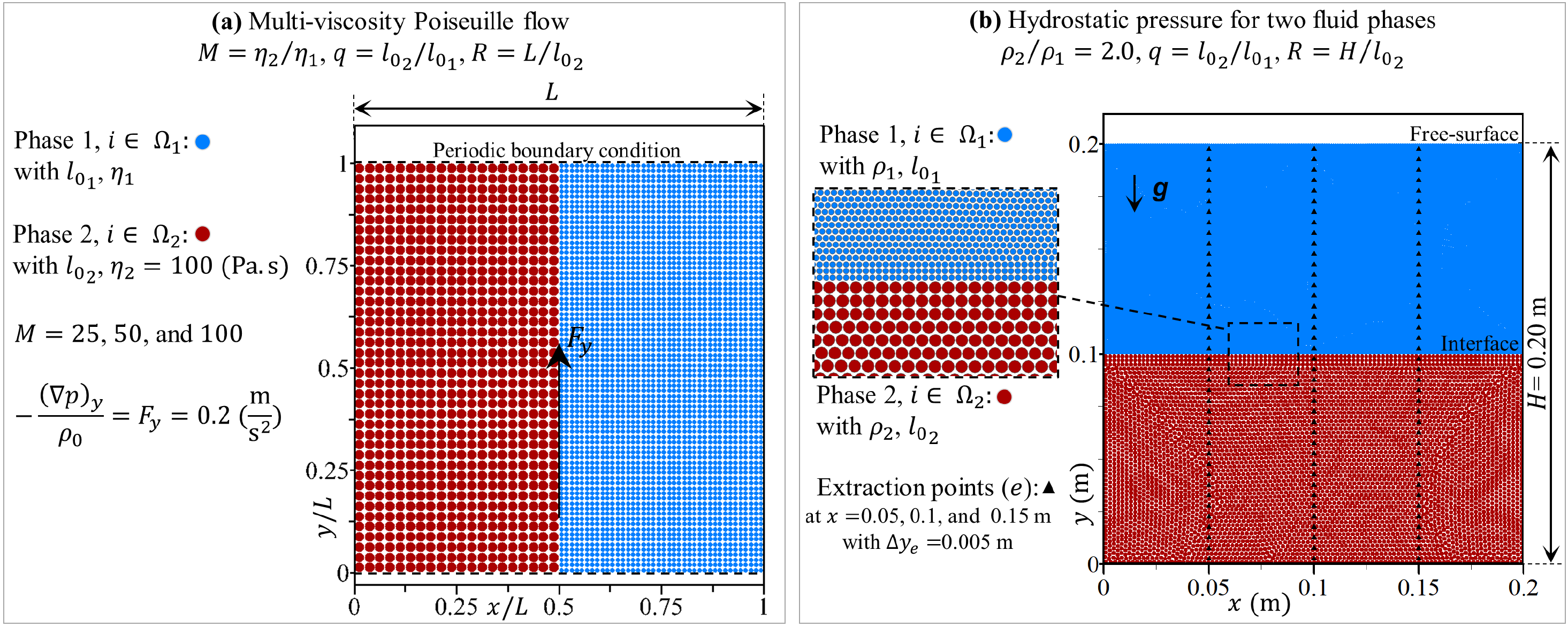}} %
	\caption{Initial configuration of the numerical simulations: (\textbf{a}) the multi-viscosity Poiseuille flow, and (\textbf{b}) the hydrostatic pressure for two fluid phases.}
	\label{fig:NCA-ini}
\end{figure}

Here, we evaluate the numerical accuracy of the multi-resolution operator in estimating the shear force by modeling the multi-viscosity Poiseuille flow. Further, we simulate the hydrostatic pressure of two fluid phases to investigate the new conservative form of governing equations in the multi-resolution configuration. Figure \ref{fig:NCA-ini} represents the initial setup of the test cases and their parameters. 

\subsubsection{Multi-viscosity Poiseuille flow}\label{sec:PF}
This benchmark case has been widely simulated for studying the numerical accuracy and convergence characteristics of particle methods \cite{Alex2018,Duan2015,Shakibaeinia_cmame_2012}. Two fluid phases, with different viscosity, flow between two stationary and parallel plates under a constant body force applied as a gradient of pressure, $ {\rho_0}{F_y}=-{\nabla p}_y$, in the positive y-direction (shown in Figure \ref{fig:NCA-ini}-a). The fluid phases, denoted as $\Omega_1$ and $\Omega_2$, fill the channel with identical width (equal to $ L/2 $) and density, i.e., $ \rho_1=\rho_2=\rho_0=1000 $ $\mathrm{ kg.m^{-3}} $. We set the viscosity ratio, $ M=\eta_2/\eta_1 $, to 25, 50, and 100 with the dynamic viscosity of the second phase, $ \eta_2 $, set to 100 Pa.s. No-slip boundary condition determines the velocity of the fixed solid boundary particles interacting with the fluid particles (i.e., $ \boldsymbol{v}_j=-\boldsymbol{v}_i $ for $ i\in\Omega_f=(\Omega_2\cup\Omega_1) $ and $ j\in\Omega_s $). Periodic boundary condition eliminates kernel truncation at the top and bottom boundaries (i.e., at $ y/L=0$ and 1). The particle size of the fluid phase with the greater viscosity (i.e, ${l_0}_2$) determines the spatial resolution of the problem by $ R=L/{l_0}_2 $. The fluid phase with the smaller viscosity value (i.e, $ \Omega_1 $) has the higher spatial resolution where the particle size ratio, defined as $ q={l_0}_2/{l_0}_1 $, is set to 2 and 4. The maximum analytical velocity, $ U_{max}$, occurs at the midpoint of $ \Omega_2 $ (i.e., at $ x/L=0.75 $) and the analytical velocity at the interface (i.e., at $ x/L=0.5 $) is denoted as $ U_0 $. The non-dimensional time, $ T $, is given by $ t{U_0}/L $ and equation (\ref{eq:DT}) determines the calculation time steps with $ C_\mathit{CFL}/c_0=0.05 $ s/m, $ C_\mathit{v}=0.25$, and $ \eta_2=100 $ Pa.s. We estimate the normalized root-mean-square-error of the velocity magnitude through $ L_2(\|\boldsymbol{v}\|)=[U_{max}]^{-1} \sqrt{1/{N_\mathit{tp}} \sum_{\forall i\in{\Omega_f}}[\|\boldsymbol{v}\|_i-\|\boldsymbol{v}\|_\mathit{analytical}^{\text{at }{x_i}}]^2} $ in which $ N_\mathit{tp} $ is the total number of fluid particles. Comparing the analytical solution of the velocity field (represented by \cite{Cao2004}) with the results of the single- and multi-resolution simulations, we investigate the numerical accuracy of the model solely related to the shear force operator.

\begin{figure}
	\centering{\includegraphics[width=\textwidth]{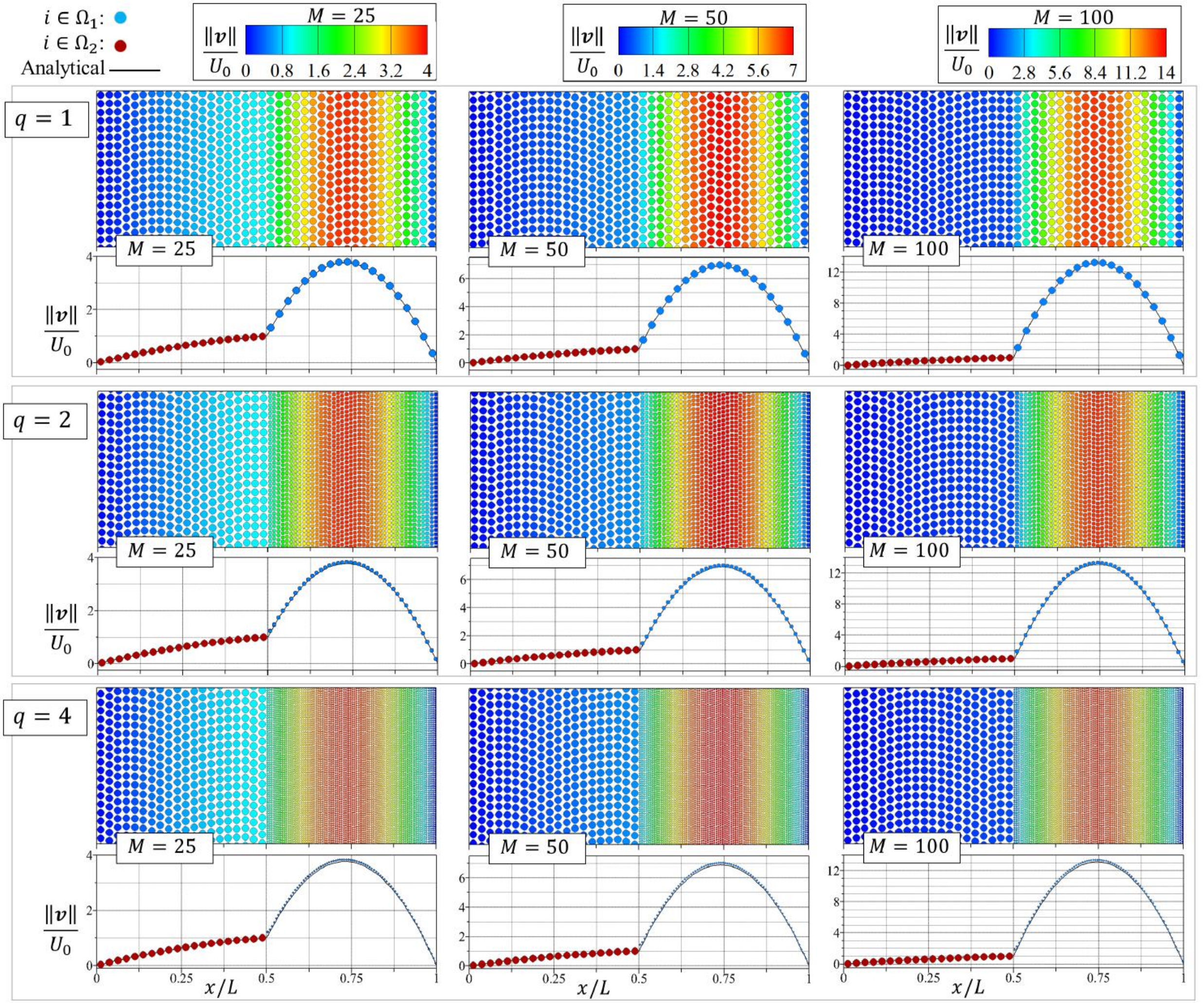}} %
	\caption{Poiseuille flow: velocity of fluid particles ($  \forall i\in \Omega_f $) at $ T=t{U_0}/L=100 $ with $ M=25 $, $ 50 $, and $ 100 $, simulated by the single-resolution ($ q={l_0}_2/{l_0}_1=1 $) and multi-resolution ($ q=2 $, $ 4 $) MPS models. The spatial resolution of the second fluid phase ($ \Omega_2 $) is $ R=L/{{l_0}_2}=40 $. The solid black lines represent the analytical velocity profiles. The  magnitude of velocity is normalized by $ U_0 $. }
	\label{fig:PF-VE}
\end{figure}

Figure \ref{fig:PF-VE} illustrates and plots the velocity of the fluid particles for different $ M $ and $ q $, at $ T=100 $ and with $ R=40 $. The single- and multi-resolution simulations predict accurate results as the estimated velocity converges to the analytical velocity profiles. With $ q=4 $, small incompatibility between the numerical and analytical results appears at the interface (i.e., at $ x/L=0.5 $) and where the maximum velocity occurs (i.e., at $ x/L=0.75 $). This discrepancy originates from the adopted assumption that considers the normalization factor (i.e., $ n_0 $) to remain valid at the interface (even where $ q\ne1 $). Also, the approximation term has no renormalization matrix for ensuring the first-order accuracy of the estimated velocity field. 

\begin{figure}
	\centering{\includegraphics[width=\textwidth]{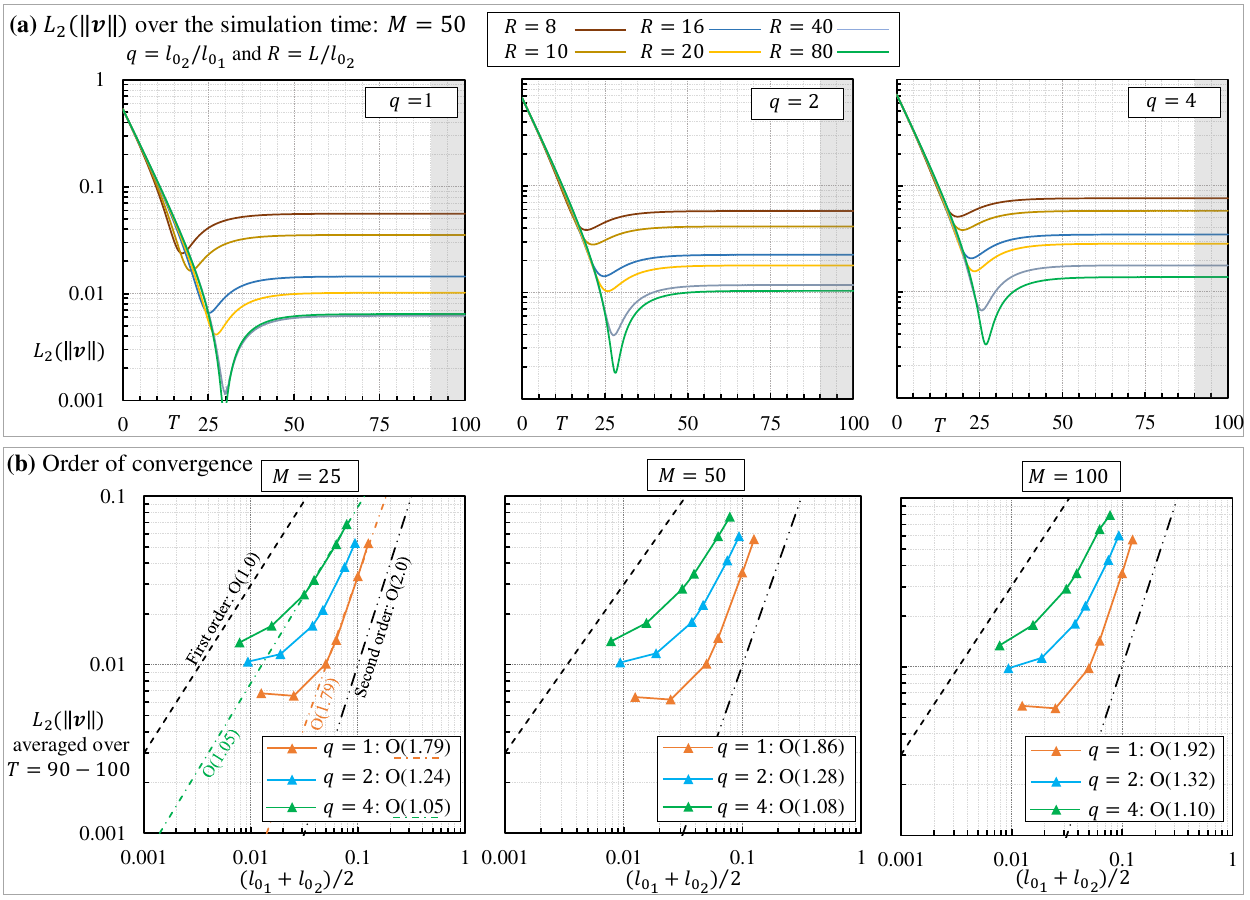}} %
	\caption{Poiseuille flow: (\textbf{a}) $ L_2(\|\boldsymbol{v}\|) $ over the simulation time for $ M=50 $ with different spatial resolutions ($ R $), and (\textbf{b}) the numerical errors and order of convergence (O(.)) for $ M=25 $, $ 50 $, and $ 100 $, simulated by the single- and multi-resolution MPS models. In (b), $ L_2(\|\boldsymbol{v}\|) $ is averaged over $ T=90-100 $ identified as the gray regions in (a).}
	\label{fig:PF-C}
\end{figure}

Next, we perform a particle convergence study of the numerical results compared with the analytical velocity profiles. We estimate and plot $L_2(\|\boldsymbol{v}\|)$ over the simulation time, $ T=0-100 $, and with different spatial resolutions (where $ R=8, 10, 16, 20, 40, $ and $ 80 $) (Figure \ref{fig:PF-C}). With both single- and multi-resolution simulations, $L_2(\|\boldsymbol{v}\|)$ reduces as the spatial resolution increases (shown in Figure \ref{fig:PF-C}-a for $ M=50 $). We plot $L_2(\|\boldsymbol{v}\|)$ against the averaged particle size (i.e., $ ({l_0}_1 +{l_0}_2)/2$) in the log-log graphs of Figure \ref{fig:PF-C}-b; the plots display that the accuracy of results is independent from the viscosity ratio $ M $ (e.g., with $ q=2 $ and $ R=40 $, $L_2(\|\boldsymbol{v}\|)$ is 1.15, 1.17, and 1.13 \%, for $ M=25, 50,$ and $ 100 $, respectively). On the other hand, the multi-resolution simulations affect the estimation of velocity profiles and the order of convergence by increasing the numerical errors (e.g., for $ M=100 $ and $ R=80 $, $L_2(\|\boldsymbol{v}\|)$ for $ q=1 $ increases from 0.5 \% to 0.9 and 1.3 \% by $ q=2 $ and $ q=4 $, respectively). However, adopting higher spatial resolutions decreases the errors with an order of convergence greater than one. Considering that the shear force calculation (i.e., $ \langle{\eta}{\nabla^2\boldsymbol{v}}\rangle $) does not benefit from any high-order approximation operators, overall, the errors by the multi-resolution implementations remain in an acceptable range (i.e.,  $L_2(\|\boldsymbol{v}\|)\leq2\%$ for $ R\ge20 $).

\subsubsection{Hydrostatic pressure}\label{sec:HP}
In this 2D benchmark case, two inviscid and immiscible fluids with identical heights fill a steady tank subjected to a constant gravitational force, ${g}=(0,-9.81 \mathrm{m/s^2})^\mathrm{t} $ \cite{Jandaghian2021_2,Rezavand2020}. The lighter phase ($ \Omega_1 $ with the density of $ \rho_1=1000 $ $ \mathrm{ kg.m^{-3}} $) is on the top of the heavier phase ($ \Omega_2 $) with the density ratio of $ \rho_2/\rho_1=2 $ (shown in Figure \ref{fig:NCA-ini}-b). The total fluid height, $ H $, and the initial particle size of phase 2, $ {l_0}_2 $, determine the spatial resolution as $ R=H/{l_0}_2 $. We set $ c_0 $, $ \rho_0 $, and $ C_\mathit{CFL} $ to $ 20 $ $ \mathrm{m/s} $, $ \rho_1 $, and $ 0.5 $, respectively. We activate the diffusive term (\ref{eq:Diff}) with $ \delta_\mathit{MPS}=0.2 $ while deactivate the DPC technique. The fluid particles ($ i\in\Omega_f=\Omega_1\cup\Omega_2 $) are initially located on a Cartesian lattice; the particle distribution of each phase is packed separately before starting the main simulations (similar to the packing algorithm proposed by \cite{Colagrossi2012}). The model assigns the initial hydrostatic pressure and the corresponding particle number density to the packed fluid particles \cite{Jandaghian2021_2}. We simulate this test case for 10 seconds where the non-dimensional time, $ T $, is given by $t\sqrt{\|{g}\|/H} $. The particle size ratio is denoted by $ q={l_0}_2/{l_0}_1 $ where $ q=1 $ and $ q=2,4 $ refer to the single- and multi-resolution simulations, respectively. To validate the numerical results (compared with the theoretical hydrostatic pressure), we extract the local pressure, $ p_e $, linearly averaged over the fluid particles within an influence radius of $ 1.5{l_0}_2 $ from the extraction points ($ e $) evenly distributed at $ x=0.05 $, $ 0.10 $, and $ 0.15 $ m with $ \Delta y_e=0.005 $ m (identified as the delta markers in figure \ref{fig:NCA-ini}-b). The normalized root-mean-square-error of the pressure is calculated by $ L_2(p)=[p_\mathit{max}]^{-1} \sqrt{1/{N_{\mathit{e}}} \sum_{e}[p_e^{\text{at }{y_e}}-p_\mathit{theoretical}^{\text{at }{y_e}}]^2} $ in which $ N_\mathit{e} $ is the total number of extraction points. The numerical error is normalized by the maximum theoretical pressure corresponding to each fluid phase, i.e., if $ y_e>H/2 $ then $ p_\mathit{max}=0.5H\rho_1\|{g}\| $ and if $ y_e<H/2 $ then $ p_\mathit{max}=0.5H(\rho_1+\rho_2)\|{g}\| $. Through this benchmark case, we investigate the accuracy of the multi-density model in predicting hydrostatic pressure. 

\begin{figure}
	\centering{\includegraphics[width=4.6 in]{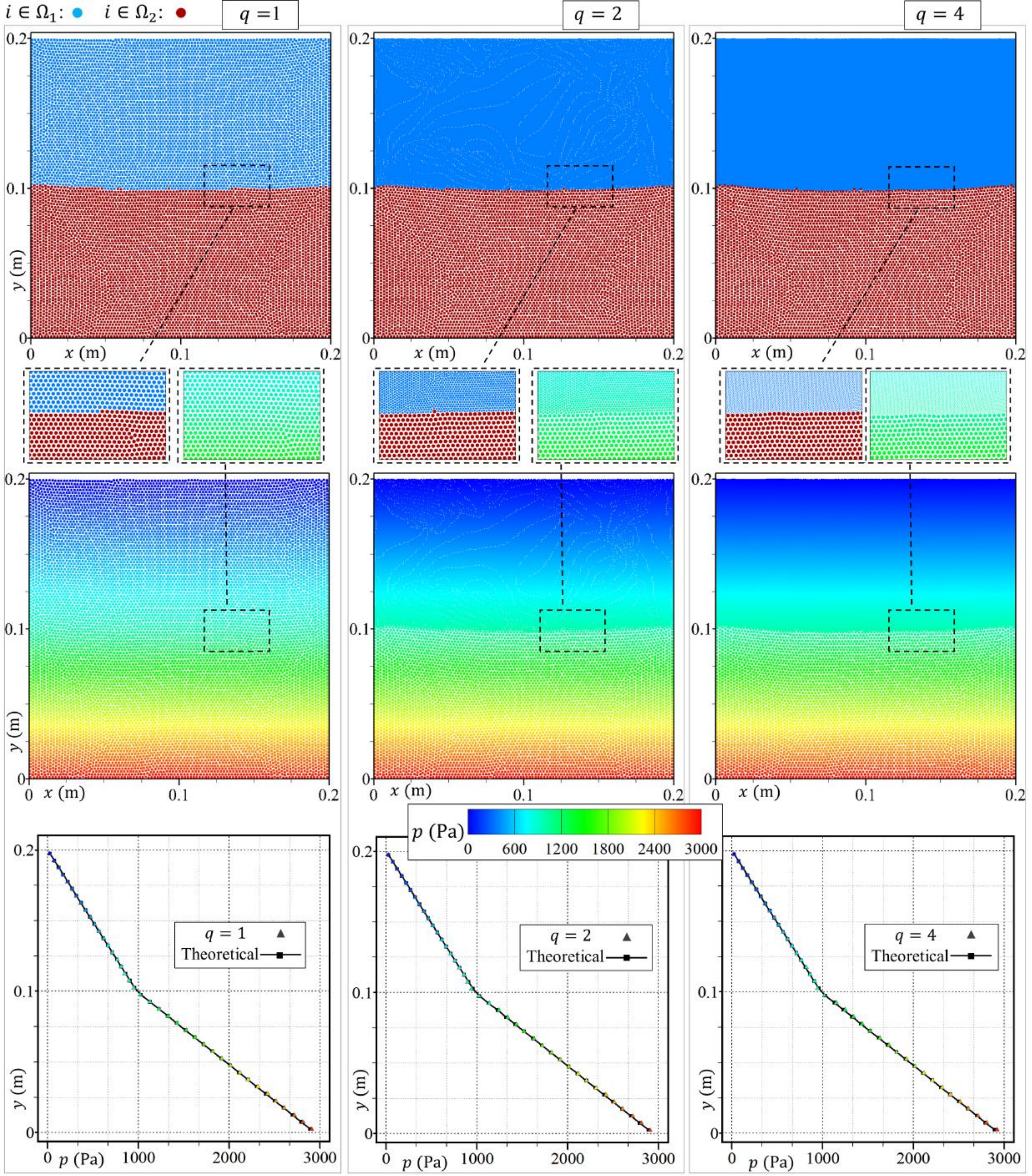}} %
	\caption{Hydrostatic pressure: particle distributions and pressure fields (with $ R=H/{l_0}_2=100$) at $ t=10 $ s (represented in the top and the middle rows, respectively). The local pressures extracted at $ x=0.1 $ m are plotted against the theoretical hydrostatic pressure (in the bottom row).}
	\label{fig:HP-RPg}
\end{figure}

Figure \ref{fig:HP-RPg} represents the particle distributions and pressure fields with $ R=100 $ and $ q=1,2$, and $4 $ at $ t=10 $ s. Stable and uniform particle distribution exists at the interface of the multi-resolution simulations (where $ q=2,4 $); the implemented diffusive term ensures smooth pressure fields over the entire fluid domain. We plot the local numerical pressures to compare with the hydrostatic pressure profile. The graphs show good agreement between the numerical results and the theoretical pressure for all three cases. 

\begin{figure}
	\centering{\includegraphics[width=\textwidth]{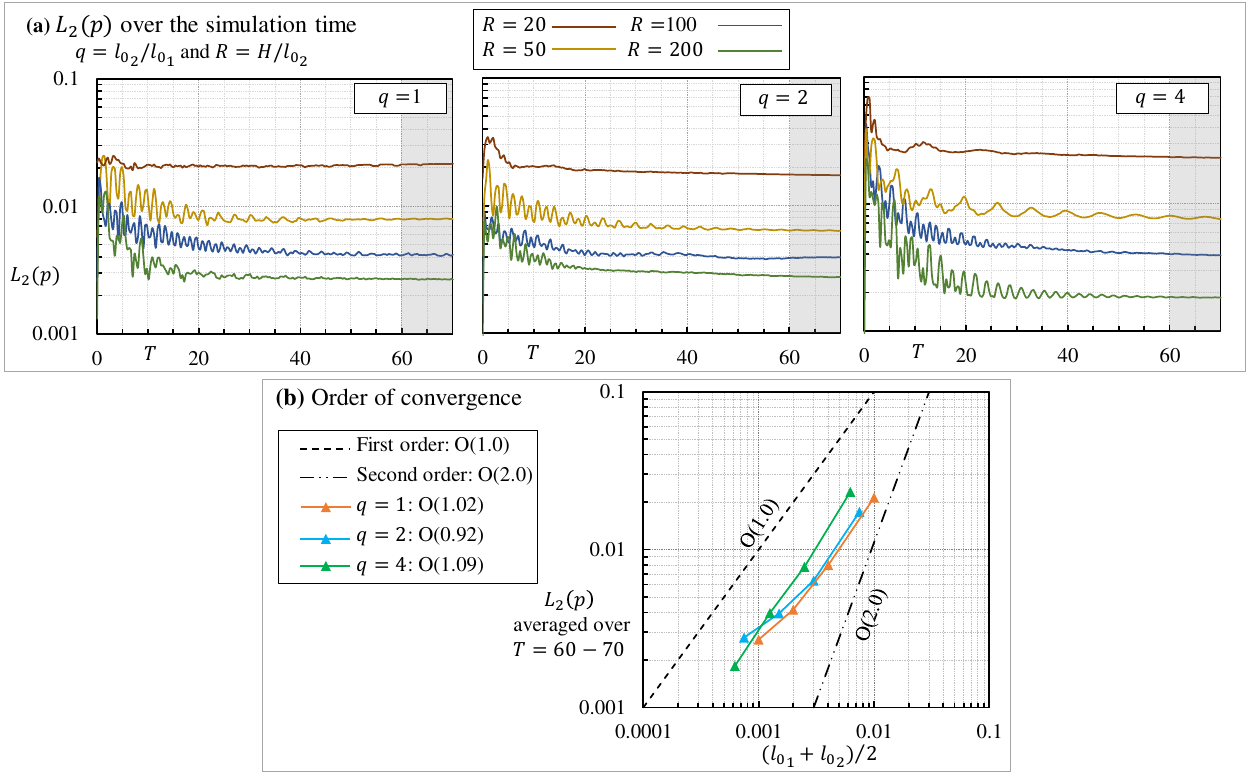}} %
	\caption{Hydrostatic pressure: (\textbf{a}) $ L_2(p) $ over the simulation time with different spatial resolutions ($ R $), and (\textbf{b}) the numerical errors and order of convergence, simulated by the single- and multi-resolution MPS models. In (b), $ L_2(p) $ is averaged over $ T=60-70 $ identified as the gray regions in (a). The local numerical pressures are averaged on the extraction points (i.e., at $ x=0.05 $, $ 0.10 $, and $ 0.15 $ m, shown in Figure \ref{fig:NCA-ini}-b).}
	\label{fig:HP-C}
\end{figure}

To quantify the accuracy and the convergence order of the results, we calculate the normalized root-mean-square-error of the pressure parameter ($ L_2(p) $) for different spatial resolutions (i.e., $ R=20,50,100 $, and $ 200 $). The numerical error is normalized by the maximum theoretical pressure corresponding to each fluid phase. The particle rearrangement (due to the assigned pressure field and the initial particle distribution at the interface) oscillates the estimated error at the initial time steps until the simulation reaches a stable condition (for $ T>40 $) (Figure \ref{fig:HP-C}-a). The numerical errors of the single- and multi-resolution simulations reduce as we increase the spatial resolution of each fluid phase. We represent $ L_2(p) $ against $ ({l_0}_1 +{l_0}_2)/2$ in a log-log plot in Figure \ref{fig:HP-C}-b. Thanks to the conservative form of the approximation operators (i.e., $ \langle{\nabla\cdot\boldsymbol{v}}\rangle $ and $ \langle\nabla{p}\rangle $) and the effective diffusive term (\ref{eq:Diff}), the accuracy of numerical results proves to be independent of $ q $ where $ R\ge100 $ (noting that $ L_2(p) $ becomes negligible, i.e., $ {L_2(p)}\leq 0.5 $ \%). Moreover, the particle convergence study confirms that the accuracy of multi-resolution simulations ($ q=2,4 $) improves at the expected rate by increasing $ R $, as the order of convergence remains equal to $ \sim 1$.  

\subsection{Dam break waves on erodible granular beds}\label{sec:EDB}
We simulate the water dam break over movable beds as a benchmark case of rapid fluid-driven granular dynamics. First, we specify the main properties of this problem and the numerical model configurations. We conduct a sensitivity analysis concerning the constant material parameters (i.e., $ \mu_2 $, $ a $, and $ b $ in the post-failure terms of the visco-inertial rheology equation) and the suspension equation (\ref{eq:sus}). We validate the results of the proposed multi-resolution MPS model against the available experimental data representing the simulated flow properties and discussing the phenomenology of the sediment erosion. Also, we evaluate the role of multi-scale water-sediment interactions within the continuum-based modeling of such multiphysics problem.

\subsubsection{Problem characteristics and configurations}\label{sec:EDB_0}
\begin{figure}
	\centering{\includegraphics[width=\textwidth]{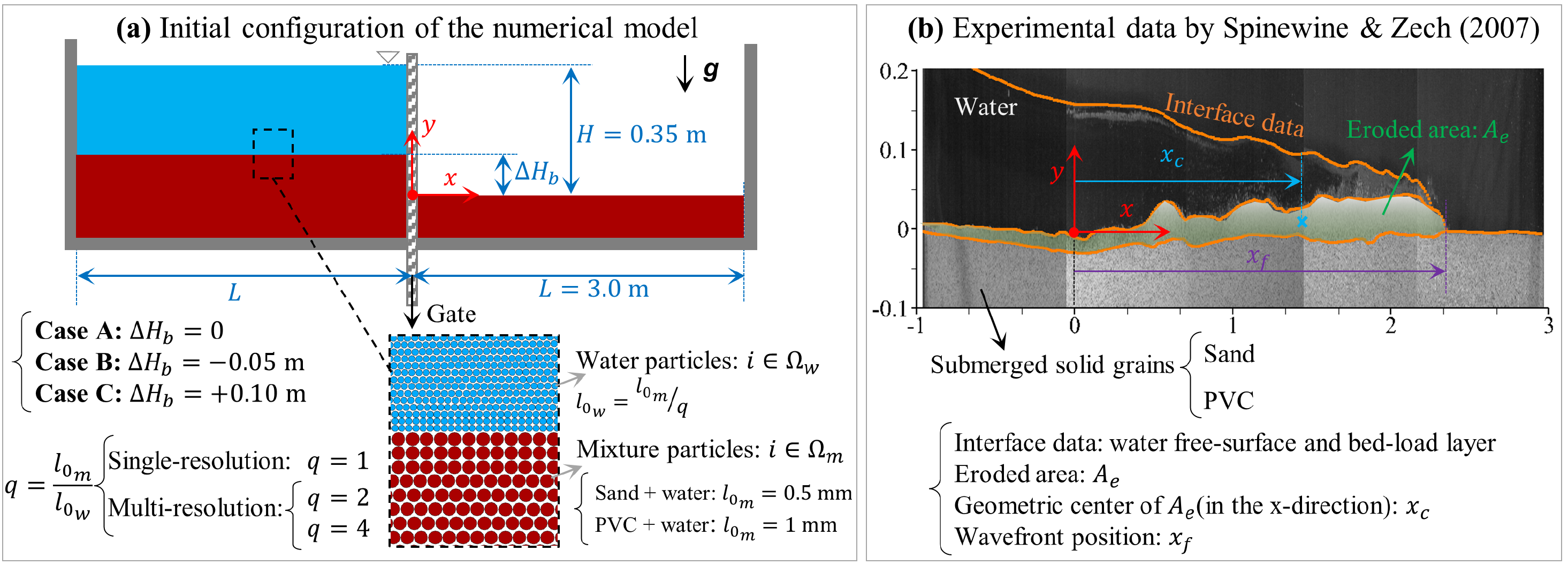}} %
	\caption{Water dam break, under the gravitational force, $ \mathbf{g}=(0,-9.81 \mathrm{m/s^2})^\mathrm{t} $, on movable beds. (\textbf{a}) The initial configuration of the two-dimensional numerical simulations, and (\textbf{b}) the experimental data by \cite{Spinewine2007}. Considering different levels of sediment on the left side of the gate, $ \Delta H_b $, we simulate cases A, B, and C (denoted as (a), (b), and (d) in the experiments, respectively). The fluid particles ($ i\in \Omega_f=\Omega_w\cup \Omega_m $) are packed before the initialization of the main simulations.}
	\label{fig:EDini}
\end{figure}

We configure the two-dimensional numerical model based on the experimental setup of \cite{Spinewine2007} (shown in Figure \ref{fig:EDini}). In this problem, a column of water collapses under the gravitational force, $\mathbf{g}=(0,-9.81 \mathrm{m/s^2})^\mathrm{t} $, on sediment beds fully submerged in water. The non-cohesive sediment material consists of either coarse sand grains or Polyvinyl Chloride (PVC) pellets \cite{Spinewine2007}; table \ref{tab:1} represents their reference material properties assigned in the rheology equations (\ref{eq:mu_m}-\ref{eq:sus}). The material constants of water are its reference density, $ {\rho_0}_{w}=1000$ $ \mathrm{ kg.m^{-3}} $, and true viscosity, $ \mu_w=0.001$ $\mathrm{Pa.s} $. The flume's length is $ 2L=6.00 $. Considering different levels of sediment on the left side of the gate, $ \Delta H_b $, different geometrical configurations exist by $ \Delta H_b=0.0 $, $ -0.05 $, and $ 0.10 $, identified as cases A, B, and C, respectively. The upstream water level with respect to the downstream sediment level, $ H $, is equal to 0.35 m and identical for all three experimental setups. The gate, located at the middle of the flume, i.e., at $ x=0.0 $, is being lowered down with the nominal speed of $ \sim 5$ m/s (in the negative y-direction) \cite{Spinewine2007}. 

In continuum-based modeling of granular material, the particle size/distance must be large enough to represent a sufficient number of grains so that the continuum assumption and hence the constitutive law remain valid \cite{Guazzelli2018}. Based on the sensitivity analysis conducted by \cite{Jandaghian2021_2} and \cite{Alex2018} for the sediment dynamics problems, we fix the initial inter-particle distance of mixture particles, $ {l_0}_m$, to $ 0.005 $ m and  $ 0.01 $ m for the sand and PVC bed materials, respectively, which correspond to $ \simeq2.7d_g $ (see Figure \ref{fig:EDini}). We define the particle size ratio, $ q $, as the ratio of $ {l_0}_m $ to the initial inter-particle distance of water particles, $ {l_0}_w $, i.e., $ q={l_0}_m/{l_0}_w $. A packed particle distribution is used for initializing the main simulations. Hydrostatic pressure determines the initial density and the effective pressure of the fluid particles at $ t=0 $ s \cite{Jandaghian2021_2}. Considering the third-order polynomial spiky kernel of \cite{Shakibaeinia_wcmps_2010} and $ k=3.1 $ for the approximation operators, the reference normalization factor $ n_0=2.2414$ (which is independent of the spatial resolution of each phase). To solve the governing equations, we set $ C_\mathit{CFL} $ and $ C_\mathit{v}$ to 0.5 and 0.125, respectively, and the reference sound speed, $ c_0 $, to 40 m/s. The diffusive term (with $ \delta_\mathit{MPS}=0.6 $) and the DPC technique are implemented within all the simulations. In this test case, the viscous force is dominant, thus, the maximum viscosity, i.e., $ \eta_\mathit{max} $ of the mixture phase, determines the time steps of the calculations through equation (\ref{eq:DT}) (which we have set to 4000 and 6000 Pa.s for the sand and PVC cases, respectively). To simulate the physical gate, we implement the virtual gate (VG) technique proposed by \cite{Jandaghian2021_2}. 

We characterize the dynamics of the immersed granular flow by the interface data (i.e., the water free-surface, the dense sediment transport layer, and the bed level), the temporal evolution of the eroded area, $ A_e $, the first moment of the eroded area, $ {x_c}{A_e} $ (where $ x_c $ is the geometric center of $ A_e $), and the wavefront position, $ x_f $ (from the experimental data provided by \cite{Spinewine2007} and \cite{Spinewine2013} identified on Figure \ref{fig:EDini}-b). We extract the numerical results at time steps identical to the experimental data before the wave reaching the end of the flume (which is at $ t=0.25 $, $ 0.5 $, $ 0.75 $, $ 1.00 $, and $ 1.25 $ seconds for the sand case and $ t=0.25 $, $ 0.5 $, $ 0.75 $, $ 1.00 $, $ 1.25 $ and $ 1.50 $ seconds for the PVC case). The non-dimensional time $ T $ is given by $t\sqrt{\|\mathbf{g}\|/H} $. We normalize $ A_e $ and $ {x_c}{A_e} $ by their corresponding reference values, i.e., $ (A_e)_\mathit{exp.}^f $ and $ ({x_c}{A_e})_\mathit{exp.}^f $ (which refer to the final data from the experiment at $ t=1.25 $ s for sand and at $ t=1.50 $ s for PVC), respectively. To detect the simulated eroded area, we employ a velocity threshold (for both water and mixture particles as $ \|\boldsymbol{v}\|_{i\in \Omega_f}\gtrsim0.25 $ \cite{Alex2018}) and a minimum volume fraction value (for water particles in the vicinity of the eroded mixture particles as $ \langle\phi\rangle_{i\in\Omega_w} /\phi_0\geq0.10-0.3$ depending on the particle size). Fluid particles that satisfy the two conditions are identified as the eroded particles ($ i\in\Omega_\mathit{ed} $). Through validating the simulated flow properties, we justify the thresholds set in the detection conditions. The numerical error of the sediment erosion parameters at $ T $ is given by $ E_r(\cdot)= [(\cdot)_\mathit{numerical}^\text{at \textit{T}}-(\cdot)_\mathit{experimental}^\text{at \textit{T}}][(\cdot)_\mathit{exp.}^f\text{ or } L]^{-1}$ normalized by the corresponding reference value. We estimate the global normalized root-mean-square-error, i.e., $ L_2(\cdot)$, by $\sqrt{1/{N_t} \sum_{1}^{N_T}[E_r(\cdot)]^2} $ where $ N_T $ is the number of calculation steps (equal to 5 and 6 for sand and PVC, respectively). 

\begin{table}
	\begin{center}
			\caption{Material properties of coarse sand grains and PVC pellets.}
		\begin{tabular}{cccccccc}
			\hline
			Sediment  & $ \rho_g $ ($ \mathrm{kg.m^{-3}} $) & $ \theta $ (degree)   &  $ d_g $ ($ \mathrm{mm} $)  & $\phi$& $ \mu_2$ &a&b \\[3pt]
			\hline
			Sand   &2683& 30 & 1.82 & 0.53 &0.84&1.23&0.3\\
			\hline
			PVC   & 1580&38 & 3.9 &0.58 &1.00&1.23&0.3\\
			\hline
		\end{tabular}

		\label{tab:1}
	\end{center}
\end{table}

\subsubsection{A sensitivity analysis of the rheology parameters}\label{sec:EDB_1}
The rheology model dynamically estimates the effective viscosity of the fluid particles as functions of the flow and material properties. \cite{Spinewine2007} reported the reference material parameters of the sand and PVC bed materials (table \ref{tab:1}); however, the constant parameters in the post-failure terms of the implemented visco-inertial model (i.e., $ \mu_2 $, $ a $, and $ b $) remain unknown and should be calibrated. Here, we analysis the sensitivity of the numerical results of case A to the rheology constants (considering the suggested values by \cite{Baumgarten&Kamrin2019}, \cite{Shi2019} and \cite{Cheng-Hsien2016}). Also, we quantify the role of the suspension equation (\ref{eq:sus}) in the overall mechanical behavior of the sediment erosion.

\begin{figure}
	\centering{\includegraphics[width=\textwidth]{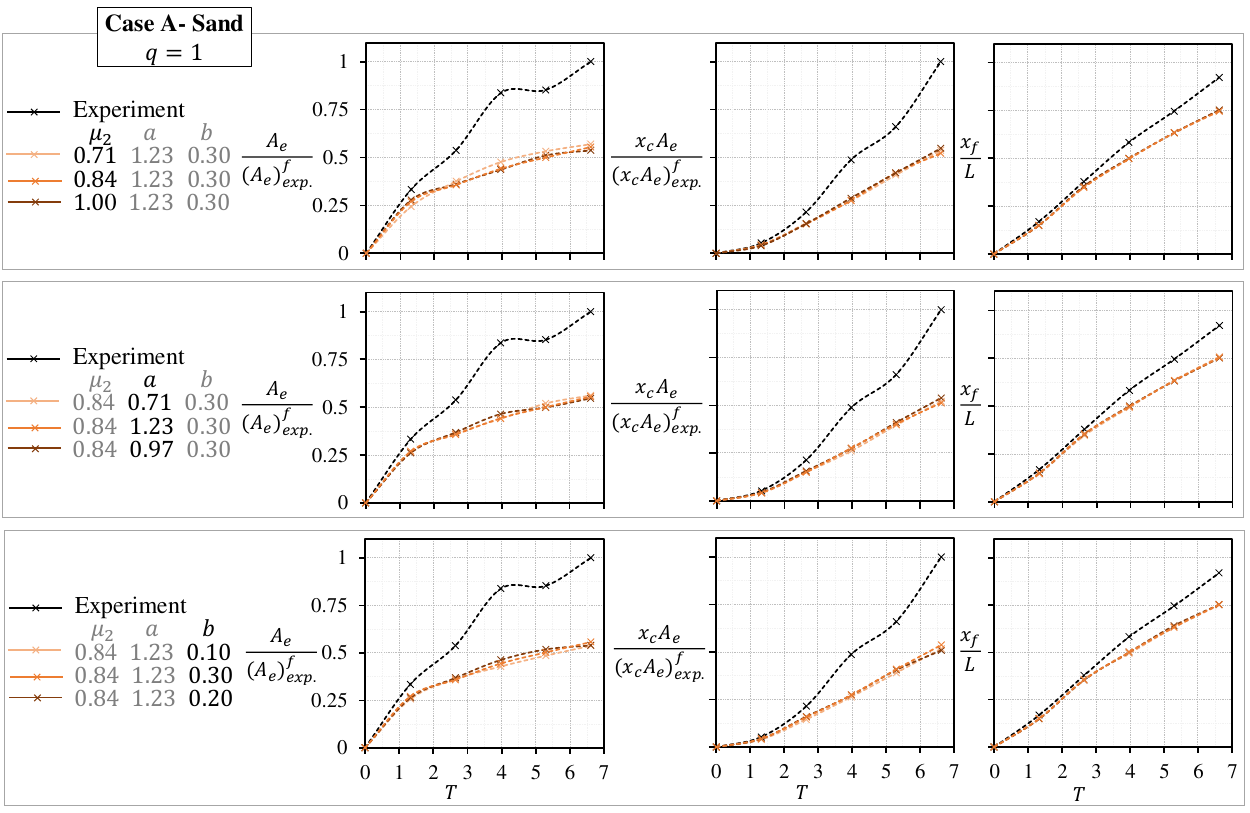}} 
	\caption{Sensitivity analysis of the fluid-driven granular flow (case A-Sand) to the rheology parameters ($ a $, $ b $, and $ \mu_2 $) through the single-resolution MPS model ($ q=1 $). The experimental profiles of the eroded area, $ A_e $, the first moment of the eroded area, $ {x_c}{A_e} $, and the wavefront position, $ x_f $, are extracted from the interface data by \cite{Spinewine2007}.}
	\label{fig:SAS}
\end{figure}

\begin{figure}
	\centering{\includegraphics[width=\textwidth]{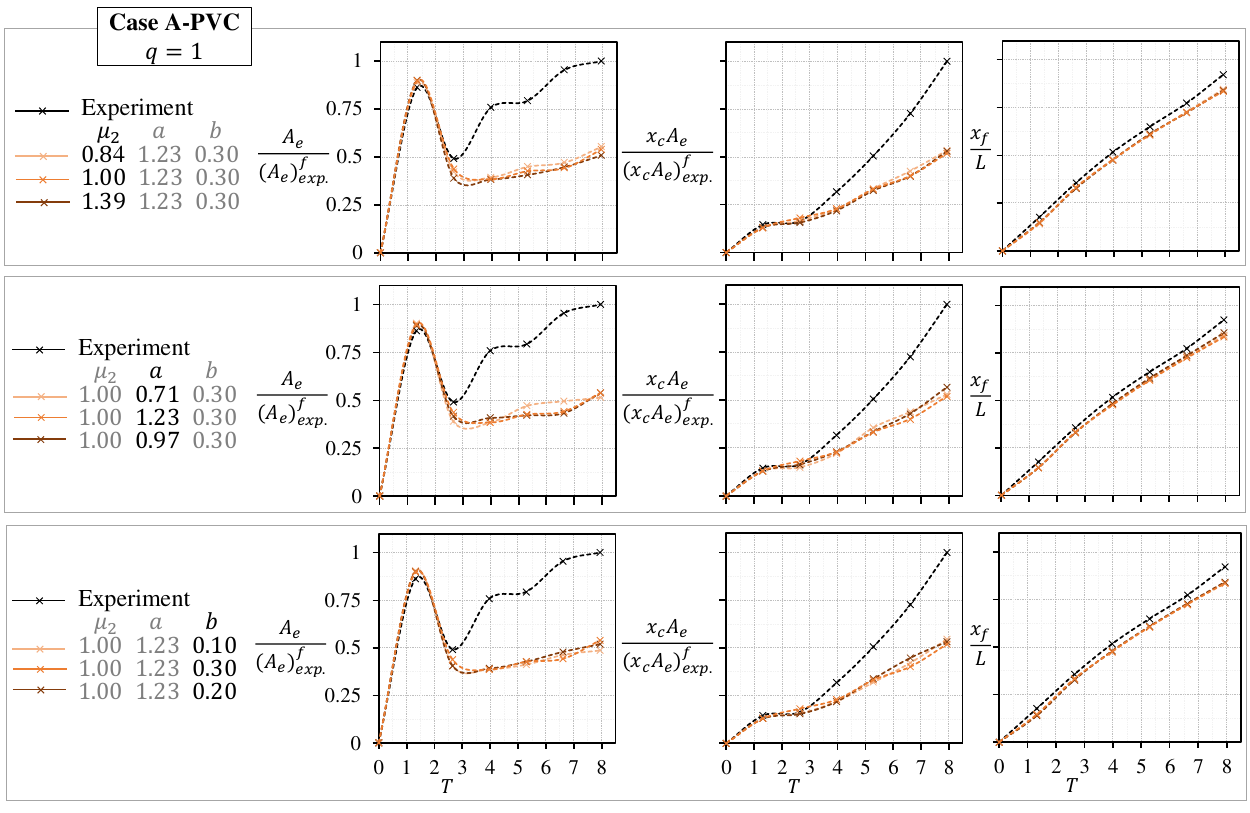}} 
	\caption{Sensitivity analysis of the fluid-driven granular flow (case A-PVC) to the rheology parameters ($ a $, $ b $, and $ \mu_2 $) through the single-resolution MPS model ($ q=1 $). The experimental profiles are extracted from the interface data by \cite{Spinewine2007}.}
	\label{fig:PAS}
\end{figure}

To conduct the sensitivity analysis, we plot the temporal evolution of $ A_e $, $ {x_c}{A_e} $, and $ x_f $ simulated by the single-resolution MPS model ($ q=1 $) where $ \mu_2 $, $ a $, and $ b $ vary as shown in Figures \ref{fig:SAS} and \ref{fig:PAS} (for the sand and PVC cases, respectively). Each parameter changes while the other two parameters are equal to their reference values given in table \ref{tab:1}. The graphs show that the bed-load evolution with different parameters of the post-failure terms remains almost alike. Table \ref{tab:4} and \ref{tab:5} represent $ L_2 $ of each scenario and compare them with the errors of the reference model (given in the first rows of the tables). For both bed materials, $ L_2 $ varies by less than $ \sim3  $\%. Thus, the sensitivity analysis confirms that the estimated sediment erosion is almost independent of the variation of the parameters, $ \mu_2 $, $ a $, and $ b $ in the specified ranges. 

Furthermore, we simulate case A-PVC (where $ q=1,2,4 $) with and without implementing the suspension equation (\ref{eq:sus}) in the rheology model. Table \ref{tab:6} shows that $ L_2(x_f) $ is almost identical for both conditions. Adding the suspension term slightly reduces $ L_2(A_e) $ and $ L_2({x_c}{A_e}) $ by $ \sim1-4 $ \%, nevertheless, the suspension term does not manipulate the overall sediment dynamics estimated by the single- and multi-resolution models. 

The considerable incompatibility between the numerical simulations and the experimental measurements (shown in the graphs of Figures \ref{fig:SAS} and \ref{fig:PAS} and quantified in table \ref{tab:6}) manifests the incapability of the single-resolution model in capturing accurate flow evolution. The continuum-based numerical model ignores some physical properties of the water-sediment mixing process associated with multi-scale interactions and volume fraction variations. No inter-particle mass exchange occurs in the numerical simulations; therefore, the model neglects the microscopic fluid flow around and between solid grains and the effects of density changes in the rheology model.

\begin{table}
	\begin{center}
			\caption{The normalized root-mean-square error ($ L_2 $) of $ A_e $, $ {x_c}{A_e} $, and $ x_f $ of case A-Sand with values for material properties in the rheology model ($ a $, $ b $, and $ \mu_2 $) and where $ q=1 $.}
		\begin{tabularx}{0.95\textwidth}{>{\centering\arraybackslash}X >{\centering\arraybackslash}X 
		>{\centering\arraybackslash}X
		ccc}
					\hline
			$ \mu_2 $&$ a $&$ b $& $ L_2(A_e) $\%   &   $L_2({x_c}{A_e})$\% & $ L_2({x_f}) $\% \\[3pt]
						\hline
			$ 0.84 $&$ 1.23 $&$ 0.3 $&  32.01 & 25.52 & 9.88\\
						\hline
			$ (0.71,1.0)$&$ 1.23 $&$ 0.3 $&  (29.99,32.39) & (24.78,25.91) & (10.01,10.20)\\
						\hline
			$ 0.84 $&$ (0.71,0.97) $&$ 0.3 $ & (31.54,31.77) & (26.63,26.60) & (9.84,9.85)\\
						\hline
			$ 0.84 $&$1.23 $&$ (0.1,0.2) $&(33.31,31.58) & (26.86,26.62) & (10.04,9.72)\\
						\hline
		\end{tabularx}

		\label{tab:4}
	\end{center}
\end{table}

\begin{table}
	\begin{center}
			\caption{The normalized root-mean-square error ($ L_2 $) of $ A_e $, $ {x_c}{A_e} $, and $ x_f $ of case A-PVC with different material properties in the rheology model ($ a $, $ b $, and $ \mu_2 $) and where $ q=1 $.}
		\begin{tabularx}{\textwidth}{>{\centering\arraybackslash}X >{\centering\arraybackslash}X 
		>{\centering\arraybackslash}X
		ccc}
					\hline
			$ \mu_2 $&$ a $&$ b $& $ L_2(A_e) $\%   &   $L_2({x_c}{A_e})$\% & $ L_2({x_f}) $\% \\[3pt]
						\hline
			$ 1.00$&$ 1.23 $&$ 0.3 $&  35.41 & 25.06 & 5.08\\			\hline
			$ (0.84,1.39)$&$ 1.23 $&$ 0.3 $&  (33.89,36.52) & (24.01,24.88) & (4.67,4.83)\\			\hline
			$ 1.00 $&$ (0.71,0.97) $&$ 0.3 $ & (34.18,35.35) & (23.67,22.90) & (4.63,3.99)\\			\hline
			$ 1.00 $&$1.23 $&$ (0.1,0.2) $   & (36.49,35.02) & (23.95,23.60) & (4.89,4.58)\\			\hline
		\end{tabularx}

		\label{tab:5}
	\end{center}
\end{table}

\begin{table}
	\begin{center}
			\caption{The normalized root-mean-square-error ($ L_2 $) of $ A_e $, $ {x_c}{A_e} $, and $ x_f $ with $ q=1 $, 2, and 4 for case A-PVC with and without the suspension equation (\ref{eq:sus}).} 
		\begin{tabularx} {\textwidth}{>{\centering\arraybackslash}X
		c c c c c c c c c} 
			\hline
			Case A-PVC & \multicolumn{3}{c}{$ L_2(A_e) $\%} & \multicolumn{3}{c}{$L_2({x_c}{A_e})$\%}& \multicolumn{3}{c}{$L_2({x_f})$\%}\\
			\hline
			$ q $ & \multicolumn{1}{c}{$1$} &\multicolumn{1}{c}{$2 $}&\multicolumn{1}{c}{$ 4 $} &\multicolumn{1}{c}{$1$} &\multicolumn{1}{c}{$ 2 $}&\multicolumn{1}{c}{$4 $} &\multicolumn{1}{c}{$1$} &\multicolumn{1}{c}{$ 2 $}&\multicolumn{1}{c}{$4 $} \\
			\hline
			With the suspension equation &  35.41 & 30.64 & 13.77& 25.06&14.76&2.23& 5.08&1.29&1.39\\
						\hline
			Without the suspension equation	& 35.83 & 32.86 & 16.05& 24.81&18.00&3.24&6.01 &2.11&1.38\\
						\hline
		\end{tabularx}

		\label{tab:6}
	\end{center}
\end{table}

\subsubsection{Flow properties and interface data}\label{sec:EDB_2}
\begin{figure}
	\centering{\includegraphics[width=\textwidth]{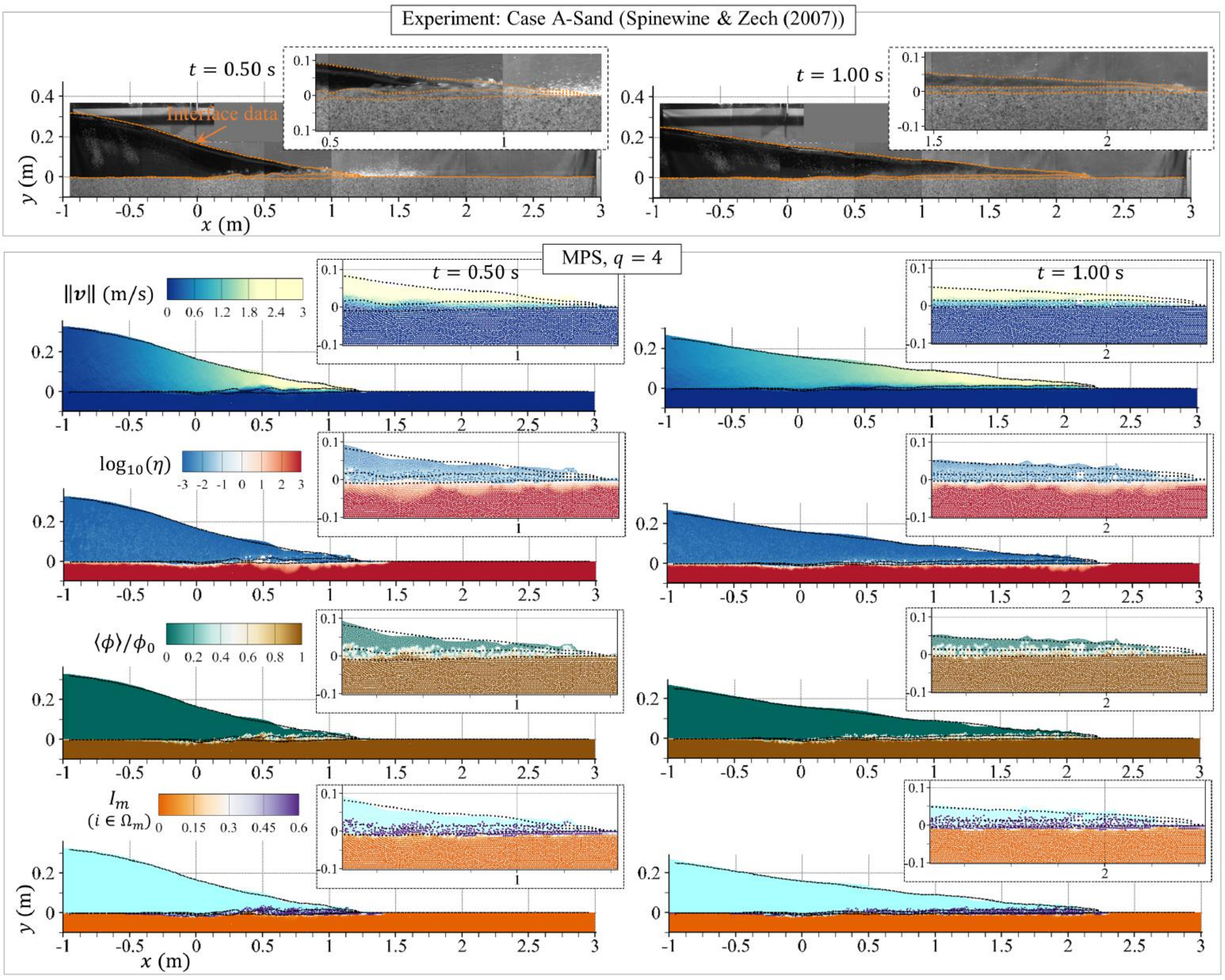}} %
	\caption{Dam break waves over the flat erodible bed (case A-Sand). The first row shows the snapshots and the interface data from the experiments by \cite{Spinewine2007}. The results of the multi-resolution MPS model (with $ q=4 $) (i.e., the velocity magnitude, $ \|\boldsymbol{v}\| $, the log of the effective viscosity, $ \text{log}_{10}(\eta)$, the approximated volume fraction, $ \langle\phi\rangle /\phi_0$, and the mixed number of the rheology model, $ I_m $) are represented at $ t=0.5 $ and $ 1.0 $ seconds. The experimental interface data (the black squares) are plotted on the numerical results. Except for the close-up plots/snapshots, the vertical scale is stretched by a factor of 1.5.}
	\label{fig:SAVMCI}
\end{figure}

\begin{figure}
	\centering{\includegraphics[width=\textwidth]{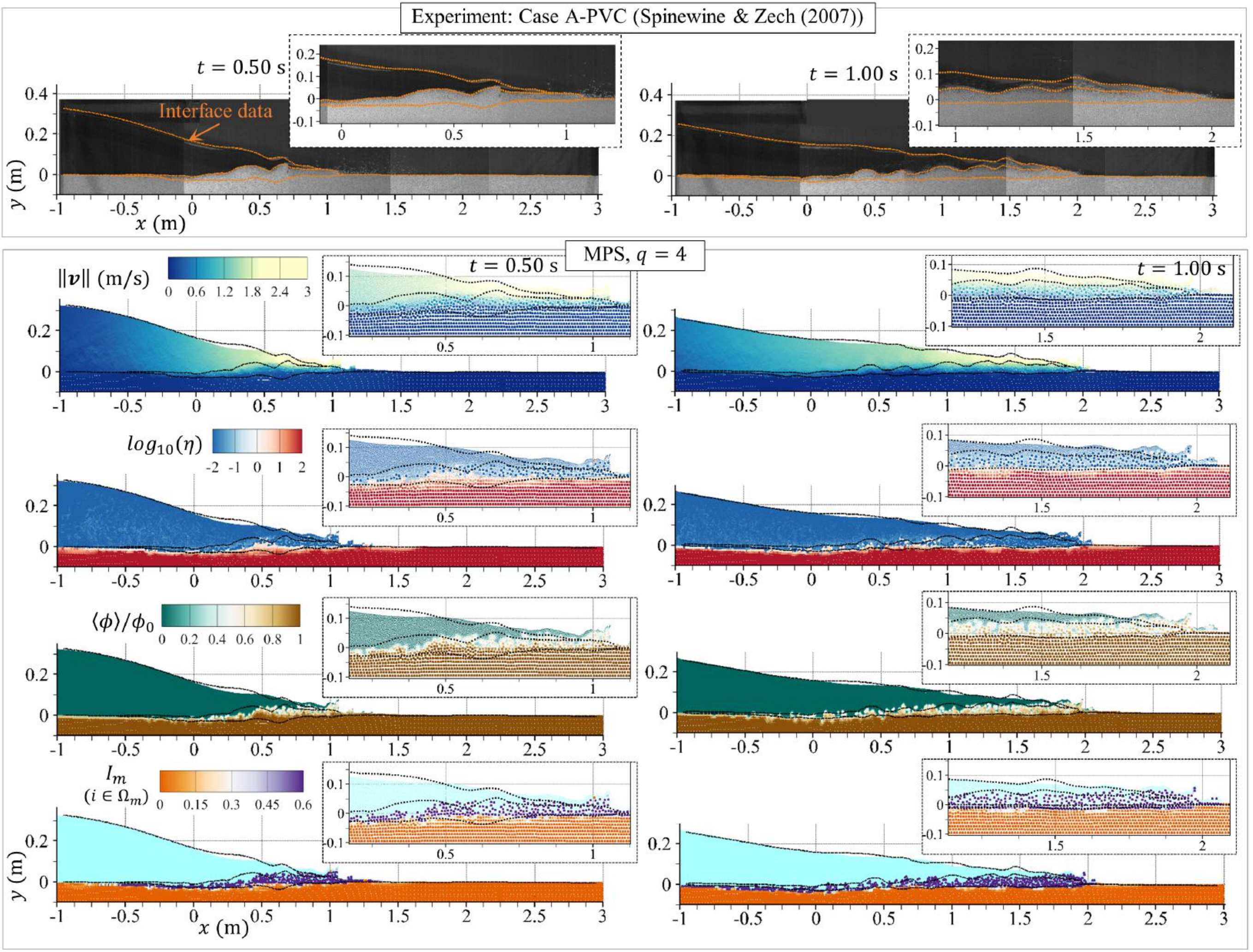}} %
	\caption{Dam break waves over the flat erodible bed (case A-PVC). The first row shows the snapshots and the interface data from the experiments by \cite{Spinewine2007}. The experimental interface data (the black squares) are plotted on the numerical results (where $ q=4 $). Except for the close-up plots/snapshots, the vertical scale is stretched by a factor of 1.5.}
	\label{fig:PAVMCI}
\end{figure}

\begin{figure}
	\centering{\includegraphics[width=\textwidth]{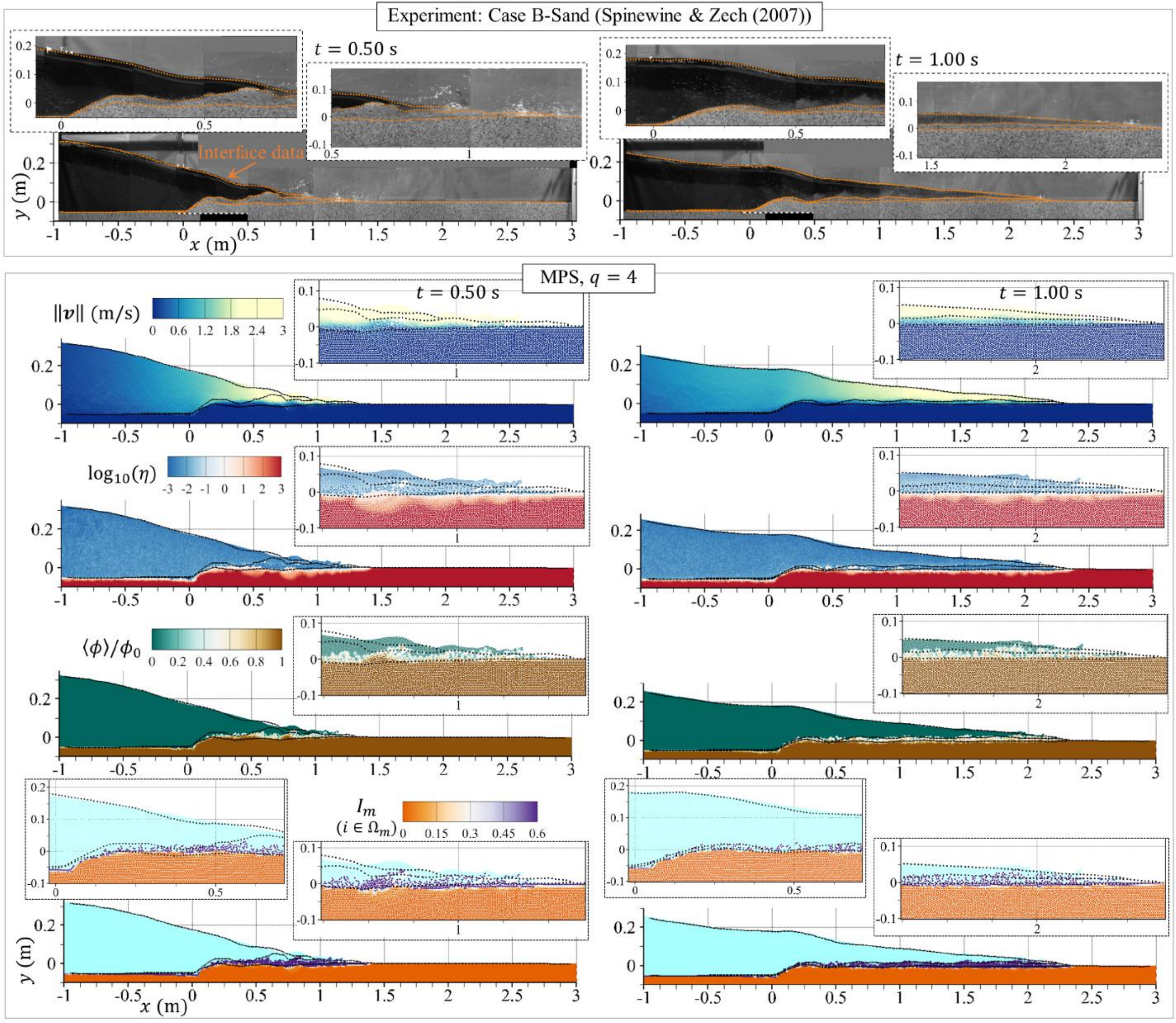}} %
	\caption{Dam break waves over the erodible bed (case B-Sand). The first row shows the snapshots and the interface data from the experiments by \cite{Spinewine2007}. The experimental interface data (the black squares) are plotted on the numerical results (where $ q=4 $). Except for the close-up plots/snapshots, the vertical scale is stretched by a factor of 1.5.}
	\label{fig:SBVMCI}
\end{figure}

\begin{figure}
	\centering{\includegraphics[width=\textwidth]{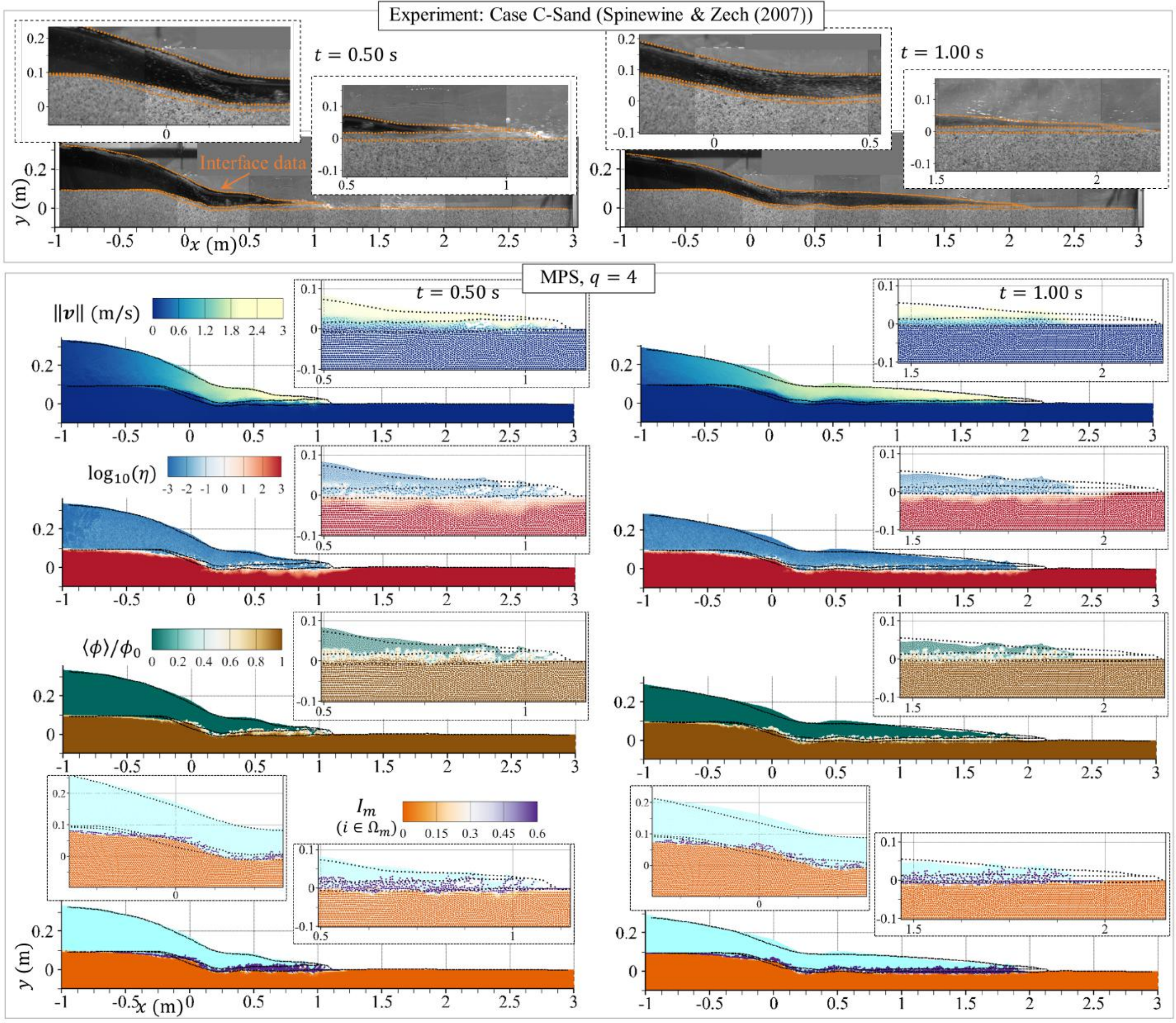}} %
	\caption{Dam break waves over the erodible bed (case C-Sand). The first row shows the snapshots and the interface data from the experiments by \cite{Spinewine2007}. The experimental interface data (the black squares) are plotted on the numerical results. Except for the close-up plots/snapshots, the vertical scale is stretched by a factor of 1.5.}
	\label{fig:SDVMCI}
\end{figure}

In this section, we present and validate the dam-break waves over erodible beds simulated by the multi-resolution MPS method (where $ q=4 $). By reporting the longitudinal and vertical flow properties, we discuss the global flow evolution and the non-linear mechanical behavior of this rapid fluid-driven problem. The velocity magnitude, $ \|\boldsymbol{v}\| $, the effective viscosity (as $ \text{log}_{10}(\eta)$), the approximated volume fractions (i.e., $ \langle\phi\rangle $ normalized by the reference volume fraction of the mixture phase, $\phi_0 $), and the mixed number, $ I_m $, of cases A-Sand, A-PVC, B-Sand, and C-Sand (at $ t=0.5 $ and $ t=1.0 $ seconds) are illustrated in Figures \ref{fig:SAVMCI}, \ref{fig:PAVMCI}, \ref{fig:SBVMCI}, and \ref{fig:SDVMCI}, respectively. The figures include snapshots of the experiments and the interface data plotted over the numerical results. Except for the close-up plots/snapshots (i.e., the inset figures with the black dash line boarder), the vertical scale of the images is stretched by a factor of 1.5 to ease visualization of the profiles and flow evolution. 

The numerical solution provides in-depth details of the water-sediment dam-break flows. As the top edge of the vertical gate reaches the bed level ($ t=0 $ s), the water column collapses on the water-saturated sediment bed driving a thin layer of bed-load toward the downstream. After the sudden vertical collapse, the wave propagates horizontally on the movable bed (considering that the wavefront position advances a distance of $ 3H $ in the positive x-direction in less than 0.5 seconds). The flow velocity increases uniformly from upstream to downstream; the dam-break surge exceeds a maximum velocity of $ 2.5 \text{ m/s}$ forming rapid erosional bores at the interface and the head of the wave (as shown in the close-up plots). For all cases the surge celerity develops similarly (independent from the initial configurations and/or the sediment materials). Furthermore, the effective viscosity field (plotted as $ \text{log}_{10}(\eta)$) illustrates the yielded and un-yielded regions estimated by the regularized rheology formulation. Inside the bed, high shear forces rapidly reduce the flow velocity towards the bottom of the flume (i.e., in the negative y-direction). The spatial variation of volume fraction at the interface manifests the mixing of water and mixture particles. From upstream to downstream, the longitudinal concentration of mixture particles increases over the bed-load layer; close to the downstream wavefront, the suspended mixture particles fill the entire flow depth (well-observed in case A with the PVC bed material) \cite{Spinewine2013}. In the implemented model, the approximated volume fraction is included in the shear force calculations through the effective viscosity terms. The mixed number of the rheology model, $ I_m $, as a function of the strain rate magnitude and the effective pressure, clearly distinguishes the suspended mixture particles (where $ I_m\gtrsim0.6 $) for which the suspension equation (\ref{eq:sus}) updates the effective viscosity. 

The overall longitudinal flow evolution, including water free-surface, the sediment transport layer thickness, and the bed boundary, are in reasonable agreement with the experimental interface data. In all cases, the high wave velocity, causing highly dynamic sediment erosion, creates irregular free-surface profiles (as rotated S-like shapes which are particularly visible at $ t=0.5 $ s). As the wave progresses on the horizontal beds (shown at $ t=1.0 $ s), the free-surface curvature reduces and better agreement exists between the numerical and experimental measurements. In the flat-bed cases (A-sand and A-PVC), the sudden vertical surge forms a scour hole at the near-dam region (i.e., the gate's location) partially captured by the numerical simulations. Also, the calculated wavefront positions of these two cases match quite well with the experimental profiles at $ t=0.5 $ and 1.0 s. In the case with a forward-facing step of the saturated sediment material (i.e., B-sand where $ \Delta H_b=-0.05 $ m), the un-yielded bed is comparable with the measured bed profile, even at the near-dam region (at $ x=0 $). However, the bed boundary of the case with the backward-facing step (i.e., C-sand where $ \Delta H_b=+0.10 $ m) does not match with the experimental profile close to the gate's location; this issue also affects the prediction of the water free-surface with stronger curvatures (at $ x=\{-0.25,+0.5\} $ m) and underestimates the wavefront position (at $ t=1.00 $ s). The observed discrepancies can be attributed to the complex non-linear flow behaviors and turbulence effects at the front of the wave which lead to non-monotonous interface profiles and non-equilibrium sediment transport \cite{Fraccarollo2002}. The continuum-based particle method struggles to accurately capture the instantaneous and local flow curvatures (especially at $ t=0.5 $ s near the wavefront of case B-PVC and the gate's location of case D-sand). We should note that the adopted numerical formulation is incapable of directly simulating the dilatation and compaction effects on the immersed granular flows. Further, the virtual gates (see \cite{Jandaghian2021_2}) ignore the gate's physical thickness, and therefore, the associated initial disturbance of its movement. Nevertheless, the developed model simulates the overall flow evolution, wave celerity, and sediment erosion processes of the dam break problem comparable with the experimental measurements and snapshots.

\begin{figure}
	\centering{\includegraphics[width=\textwidth]{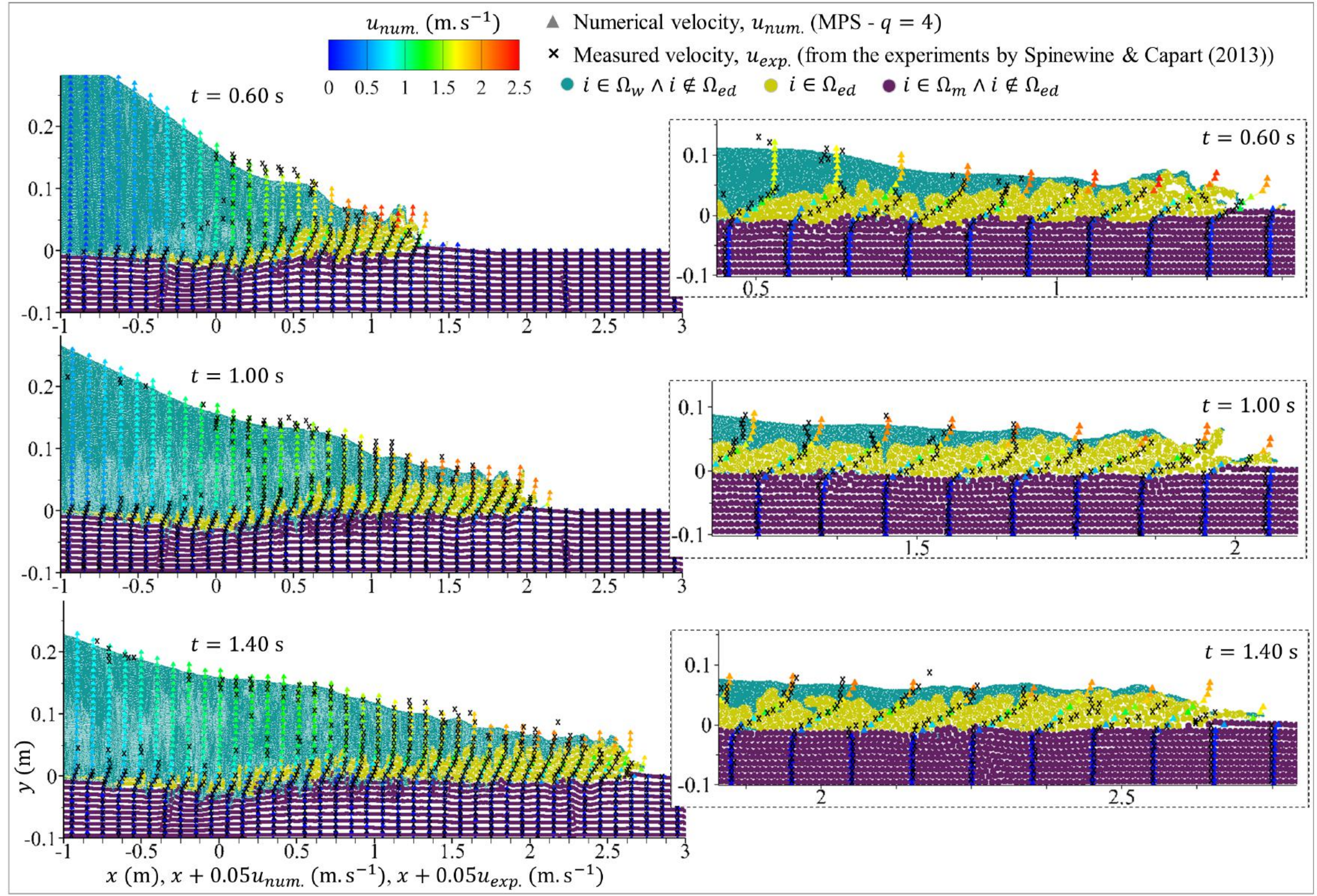}} %
	\caption{Vertical velocity profiles of dam break waves over the flat erodible bed (case A-PVC) in comparison with the experimental data at $ t=0.6 $, $ 1.0 $, and $ 1.4 $ seconds. The local numerical velocity, $ x+0.05u_\mathit{num.} $, is identified by the delta markers and the color map. The experimental data, $ x+0.05u_\mathit{exp.} $ (shown as the black x markers) are extracted from the work of \cite{Spinewine2013}. Except for the close-up plots, the vertical scale is stretched by a factor of 4.}
	\label{fig:PA-V}
\end{figure}

\begin{figure}
	\centering{\includegraphics[width=\textwidth]{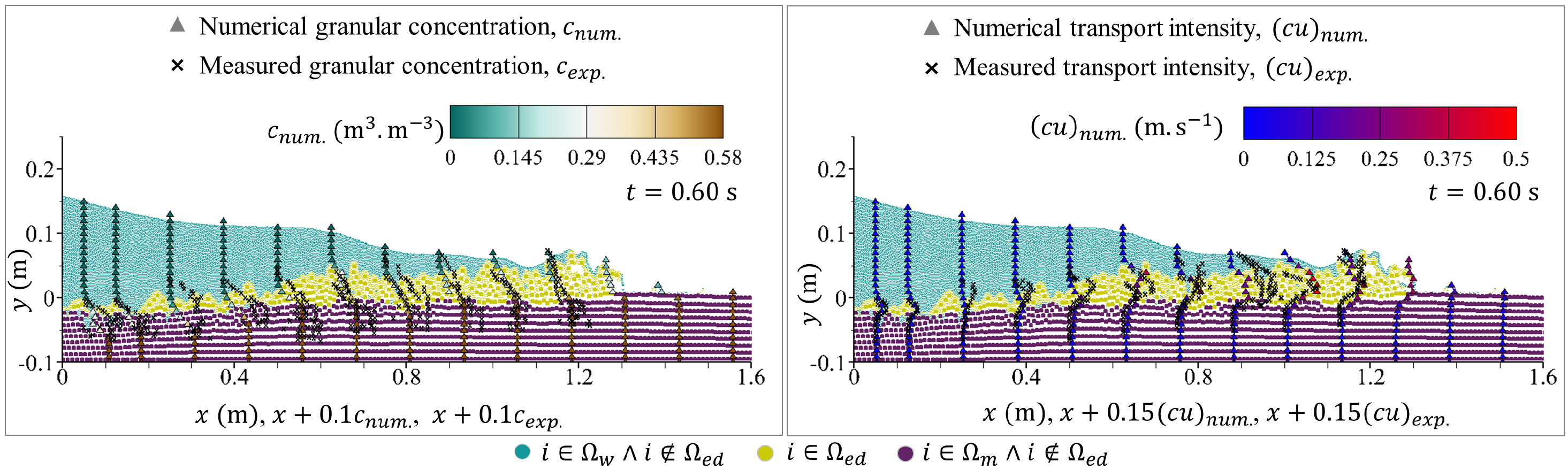}} %
	\caption{Vertical profiles of the granular concentration, $ c $, and sediment transport intensity, $ cu $, of case A-PVC at $ t=0.6 $ seconds (on the left and right graphs, respectively). The numerical results of the multi-resolution model ($ q=4 $) are compared with the experimental measurements by \cite{Spinewine2013} (plotted as the delta and the black x markers, respectively). The vertical scale is stretched by a factor of 1.5.}
	\label{fig:PA-CV}
\end{figure}

Next, we validate the internal flow properties of the numerical simulation of case A-PVC against the available experimental measurements represented by \cite{Spinewine2013} (Figures \ref{fig:PA-V} and \ref{fig:PA-CV}). \cite{Spinewine2013} used a particle tracking analysis to measure the vertical velocity profiles, $ u_\mathit{exp.} $, at 40 cross-sections evenly distributed between $ x=-0.95 $ and $ +2.95$ m with $ \Delta x=0.10 $ m. We extract the numerical velocity, $ u_\mathit{num.}  $, by linearly averaging the velocity magnitude of fluid particles at the vicinity of the extraction points with $ \Delta y=0.01 $ m on the same vertical cross-sections. Figure \ref{fig:PA-V} plots the fluid particles (classified as the eroded and not eroded particles, i.e., $ i\in \Omega_\mathit{ed} $ and  $ i\notin \Omega_\mathit{ed} $, respectively), the extracted numerical velocity as $x+0.05u_\mathit{num.} $, and the experimental velocity as $ x+0.05u_\mathit{exp.} $, at $ t=0.6 $, 1.0, and 1.4 seconds. The numerical model simulates the non-linear velocity profiles with smooth variations at the top and base of the sediment transport layer; the velocity profiles in the water layer remain uniform with maximum magnitudes at the free-surface. At $ t=1.4 $ s, the model slightly underestimates the velocity near the wavefront ($ x\simeq2.5 $). Nevertheless, the estimated velocity matches the experimental profiles quite well over the entire fluid domain. Moreover, we extract and compare the local granular concentration, $ c $, and the sediment transport intensity defined as the product of $ c $ and $ u $ (i.e., $ c$ and $cu$, respectively). \cite{Spinewine2013} reported the corresponding experimental measurements limited to the granular layer at 13 laser-instrumented vertical cross-sections between $ x=0.05 $ and $ 1.5$ m. To estimate $ c_\mathit{num.} $, we linearly average the approximated volume fraction of the particles, $ \langle\phi\rangle_{i\in\Omega_f}$, at the extraction points. One should note that the experimental measurements are missing at $ x\ge1.25 $ m and $ t=0.6 $ s (unlike the velocity profiles reported by the particle tracking analysis in Figure \ref{fig:PA-V}). We plot $x+0.1 c$ and $x+0.15cu$ over the fluid particles in the left and right graphs of Figure \ref{fig:PA-CV}, respectively. The granular concentration keeps its maximum value ($ \phi_0 $) in the granular layer while reducing across the bed-load layer toward the free surface. At the top interface of the transport layer, the granular concentration drops and further vanishes in the absence of the mixture particles. In spite of some minor discrepancies, the estimated numerical profiles are in good agreement with the experimental measurements; the validation confirms that the numerical model can predict the non-linear profiles with smooth variations across the bed-load layer. We should also highlight that the compatibility between the numerical and experimental profiles justifies conditions we have adopted to detect the eroded area (through which we study the phenomenology of sediment erosion and quantify the numerical validations). 

\subsubsection{Phenomenology of the sediment erosion: $ A_e $, $ {x_c}{A_e} $, and $ x_f $}\label{sec:EDB_3}
\begin{figure}
	\centering{\includegraphics[width=\textwidth]{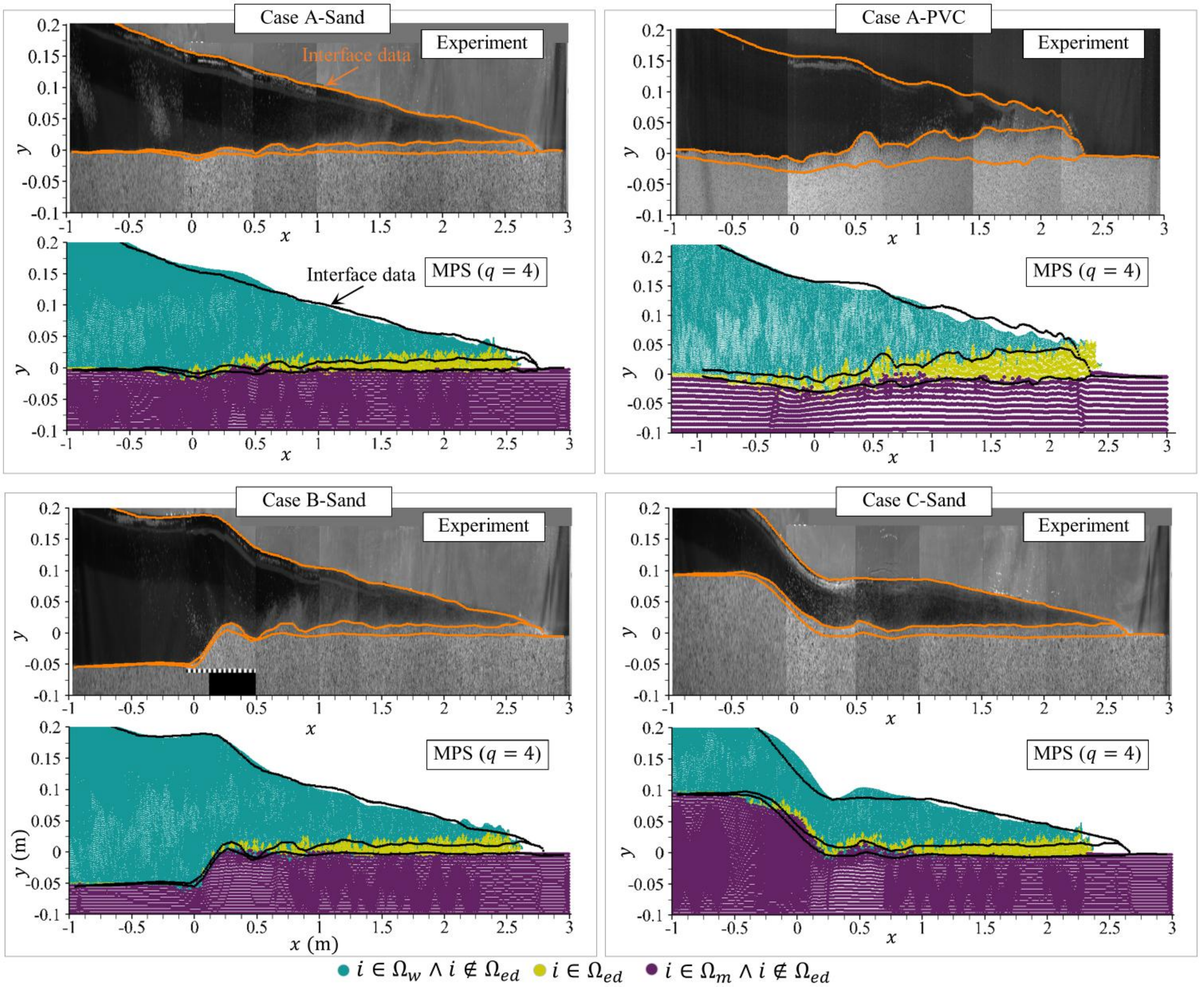}} 
	\caption{Flow evolution and the eroded region, $ i\in \Omega_\mathit{ed} $, of case A, B, and C, at $ t=1.25 $ s, simulated by the multi-resolution MPS model ($ q=4 $) compared with the experimental snapshots and interface data (the black squares) from the work of \cite{Spinewine2007}. The vertical scale is stretched by a factor of 5.0.}
	\label{fig:EDALL}
\end{figure}

\begin{figure}
	\centering{\includegraphics[width=\textwidth]{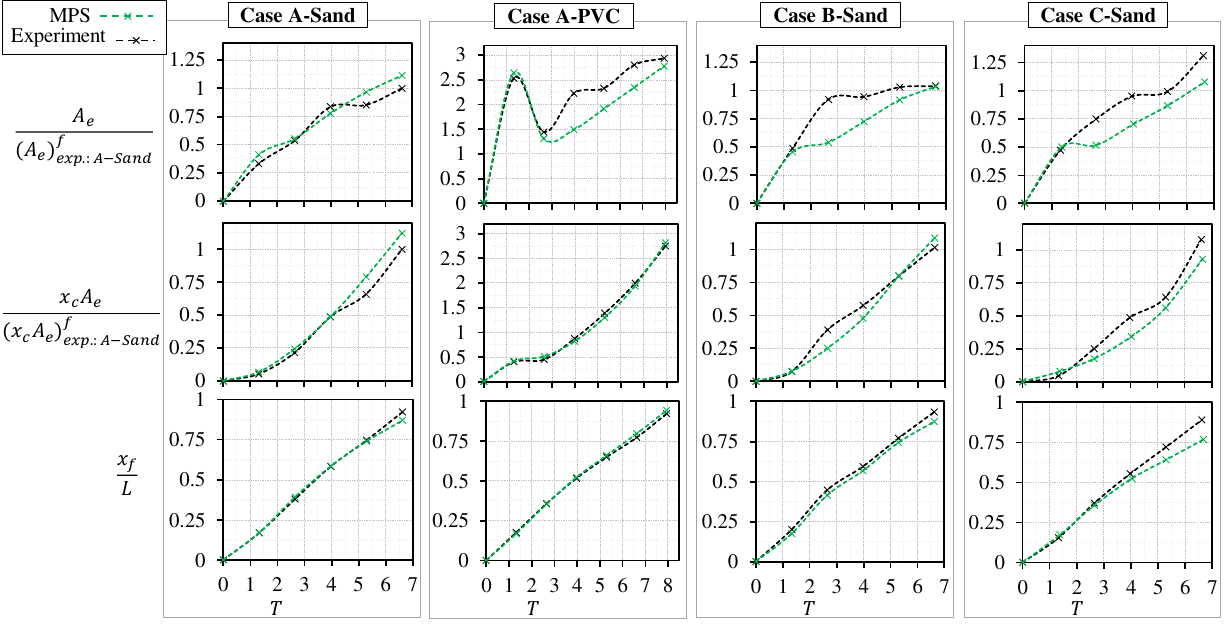}} %
	\caption{Temporal evolution of $ A_e $, $ {x_c}{A_e} $, and $ x_f $ of cases A-Sand, A-PVC, B-Sand, and D-Sand with the multi-resolution MPS model ($ q=4 $) compared with the experimental profiles (extracted from the interface data reported by \cite{Spinewine2007}). $ A_e $ and $ {x_c}{A_e} $ are normalized by the corresponding reference values of case A-Sand (denoted as $ (A_e)^f_\mathit{exp.:A-Sand} $ and $ ({x_c}{A_e})^f_\mathit{exp.:A-Sand}$, respectively). }
	\label{fig:ALLG}
\end{figure}

To overview the phenomenology of the dam break waves over movable beds, we compare the sediment dynamics of the different configurations. Figure \ref{fig:EDALL} shows the wave propagation of cases A-sand, A-PVC, B-sand, and C-sand on the horizontal beds at $ t=1.25 $ s. The figure contains experimental snapshots and the interface data of the test cases. Further, we plot the temporal evolution of the global erosion variables, $ A_e $ and $ {x_c}{A_e} $, and the wavefront position, $ x_f $, in Figure \ref{fig:ALLG}, and represent the associated numerical errors in table \ref{tab:7}.

Qualitatively, the numerical simulations provide comparable flow evolution and predict the thickness of the transport layer at the center of the wave in all four cases. The base of the transport layer is well-captured; however, near the gate's location, the rapid flow involves complex interface deformations affecting the numerical predictions. As discussed earlier, this incompatibility is more noticeable for case C-sand with the backward-facing step. Comparing case A-PVC with the other configurations illustrates the sensitivity of sediment erosion to the density ratio (i.e., $ \rho_m/\rho_w $). The thickness of the transport layer increases significantly for case A-PVC as the wave can mobilize more mixture particles. On the other hand, with the sand material (i.e., with the heavier grains, but with a smaller fiction angle) in cases A, B, and C, the transport layer is thinner; thus, the wave can progress further on the horizontal bed with less sediment erosion toward the downstream \cite{Spinewine2013, Fraccarollo2002}. While the wavefront position of case A-PVC matches well with the interface data, the numerical simulation slightly underestimates the wave propagation speed of cases with the sand beds (regardless of their initial configuration). 

To quantitatively study the sediment dynamics, Figure \ref{fig:ALLG} provides an overall overview of the sediment erosion concerning different configurations and bed materials. The graphs of Figure \ref{fig:ALLG} show that the multi-resolution MPS model predicts the global behavior of the sediment erosion and wave prorogation. Particularly, we observe that $ {x_c}{A_e} $, and $ x_f $ of all four cases are in very good agreement with the experimental data. For the light PVC material, sediment erosion increases significantly as $ A_e $ of A-PVC is almost 2.5 times greater than $ A_e $ of A-sand (at $ t=1.25 $ s). For case C-sand, the wave interaction with the backward-facing step increases the sediment erosion by $\sim30 $ percents in comparison with cases A-sand and B-sand. With the step-like discontinuity of cases B and C, the results are less satisfactory considering that the numerical model underestimates the sediment erosion at $ t=0.5$ and 0.75 s. Table \ref{tab:7} represents the numerical errors of the sediment erosion parameters, $ L_2 $. $ L_2(x_f) $ and  $ L_2({x_c}{A_e}) $ remain less than 10 percents manifesting the acceptable accuracy of numerical model in simulating the flow evolution on the movable beds. The results show that with the step-like discontinuities, $ L_2(A_e) $ increases by $ \sim10 $ percents (from $ \sim8.5 $ for case A-sand to $ \sim19.7$ and $\sim14.7$ percents). We attribute the numerical errors to the neglected role of granular dilatation and compaction in the adopted single-phase rheology model. This issue indirectly ignores the turbulence suspension effects of the microscopic pore water flow between solid grains within the numerical simulations \cite{Spinewine2013}. Nevertheless, the multi-resolution MPS model proves to be capable of predicting the global behavior of the fluid-driven granular dynamics with reasonable numerical accuracy.

\begin{table}
	\begin{center}
			\caption{The normalized root-mean-square error ($ L_2 $) of $ A_e $, $ {x_c}{A_e} $, and $ x_f $ of cases A-Sand, A-PVC, B-Sand, and C-Sand simulated by the multi-resolution MPS model ($ q=4 $).}
		\begin{tabular}{cccc}
			\hline
			Case & $ L_2(A_e) $\% &$L_2({x_c}{A_e})$\% & $ L_2({x_f}) $\% \\[3pt]
			\hline
			A-Sand& 8.49  & 8.20 & 2.40\\
			\hline
			A-PVC& 13.77 & 2.23 & 1.39\\
			\hline
			B-Sand& 19.66 & 8.09 & 3.67\\
			\hline
			C-Sand& 14.68 & 8.54 & 6.79\\
			\hline
		\end{tabular}
		\label{tab:7}
	\end{center}
\end{table}

\subsubsection{Role of multi-scale simulations}\label{sec:EDB_4}
\begin{figure}
	\centering{\includegraphics[width=\textwidth]{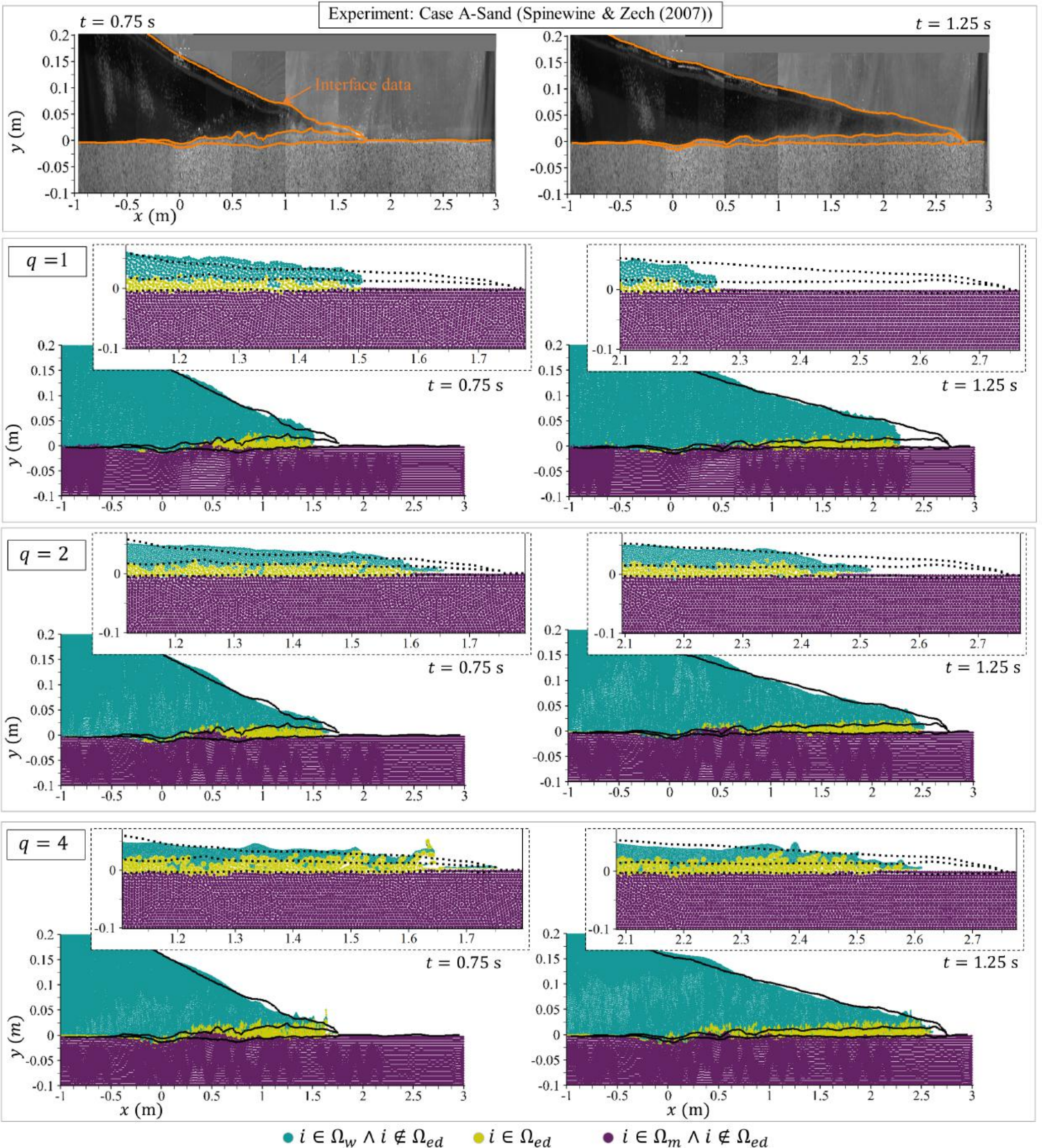}} 
	\caption{Flow evolution and the detected eroded region, $ i\in \Omega_\mathit{ed} $, for case A-sand with the single-resolution ($ q=1 $) and multi-resolution ($ q=2$, $ 4 $) MPS models compared with the experimental interface data (the black squares) from \cite{Spinewine2007}. Except for the close-up plots, the vertical scale is stretched by a factor of 5.0.}
	\label{fig:SAED}
\end{figure}

\begin{figure}
	\centering{\includegraphics[width=\textwidth]{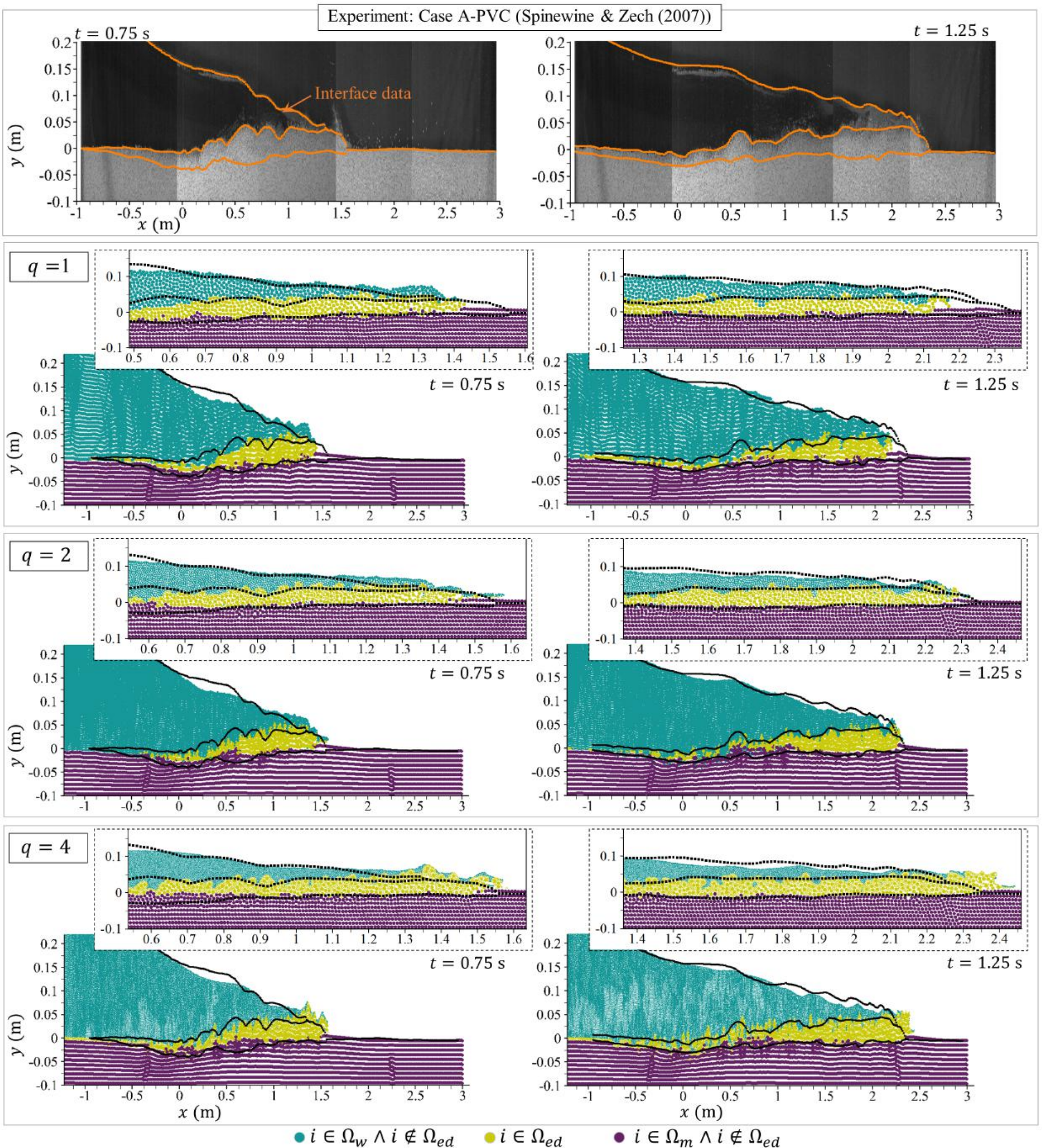}} 
	\caption{Flow evolution and the detected eroded region, $ i\in \Omega_\mathit{ed} $, for case A-PVC with the single-resolution ($ q=1 $) and multi-resolution ($ q=2$, $ 4 $) MPS models compared with the experimental interface data (the black squares) from \cite{Spinewine2007}. Except for the close-up plots, the vertical scale is stretched by a factor of 5.0.}
	\label{fig:PAED}
\end{figure}

Here, we discuss the role of multi-scale interactions in sediment erosion modeling. To do so, we compare the interface data of case A simulated by the single- and multi-resolution models. Figures \ref{fig:SAED} and \ref{fig:PAED} represent the eroded region ($ i\in \Omega_\mathit{ed} $) of case A with the sand and PVC bed materials, respectively. In these figures, except for the close-up plots, the vertical scale is stretched by a factor of 5.0. Flow evolution of the water and mixture particles (at $ t=0.75 $ and $ 1.25 $ seconds) shows that the single-resolution model ($ q=1 $) underestimates the wavefront position. This discrepancy between the numerical and experimental results is more noticeable for the sand case with a higher density ratio. On the other hand, the multi-resolution models ($ q=2,4 $) predict more flow deformations at the interface, increasing erosion of the mixture particles. Although the continuum-based modeling still misses some physical aspects of sediment erosion (i.e., the pore water flow between the solid grains and the associated changes in the volume fraction of the mixture particles), the multi-resolution particle interactions allow the numerical model to capture more accurate flow evolution related to the multi-scale feature of water-sediment dynamics. 

\begin{figure}
	\centering{\includegraphics[width=\textwidth]{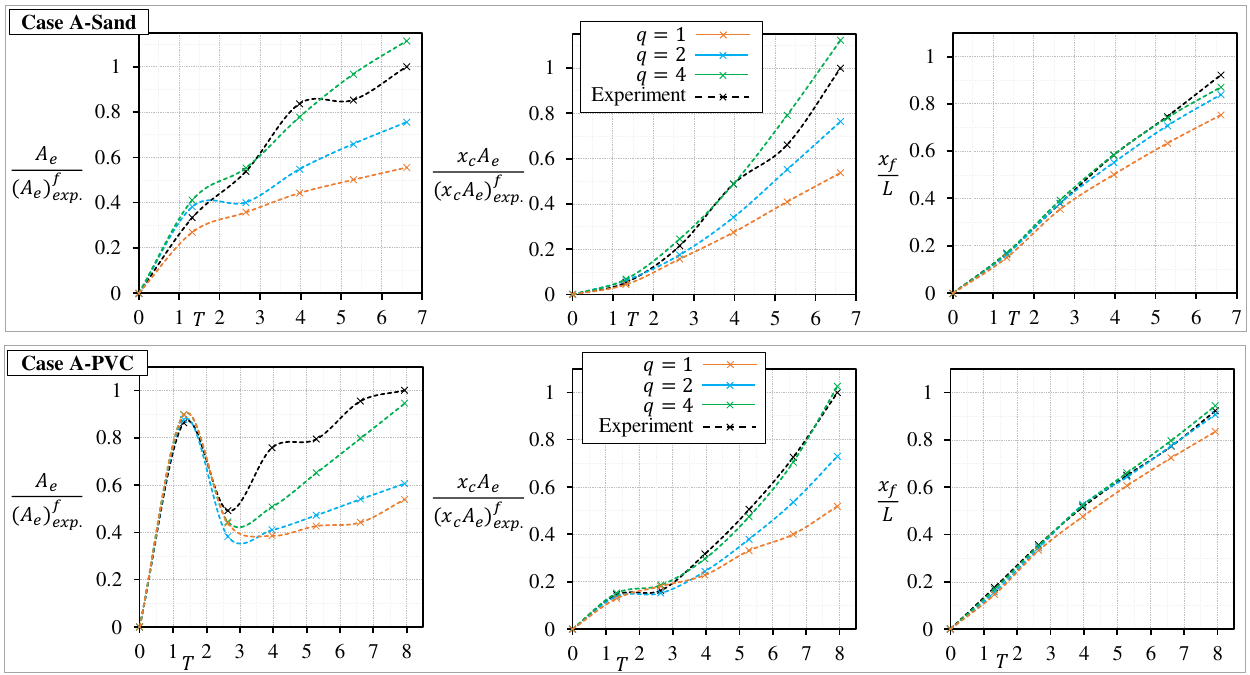}} 
	\caption{Temporal evolution of the eroded area, $ A_e $, the first moment of the eroded area, $ {x_c}{A_e} $, and the wavefront position, $ x_f $, with the single-resolution ($ q=1 $) and multi-resolution ($ q=2$, $ 4 $) MPS models for case A with sand and PVC compared with the experimental measurements by \cite{Spinewine2007}.}
	\label{fig:SAPA}
\end{figure}

\begin{figure}
	\centering{\includegraphics[width=12 cm]{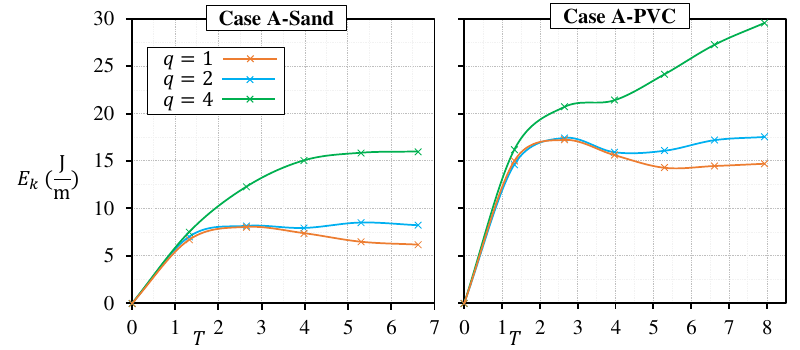}} 
	\caption{The global kinetic energy of the mixture particles ($ E_k=0.5\sum {\rho_{0}}_i{{l_0}_i}^2{\|\boldsymbol{v}\|_i}^2 $ for $ \forall i\in \Omega_m $) with the single-resolution ($ q=1 $) and multi-resolution ($ q=2$, $ 4 $) MPS models for case A with sand and PVC.}
	\label{fig:Ek}
\end{figure}

To quantify the numerical validations, we plot the temporal evolution of $ A_e $, $ {x_c}{A_e} $, and $ x_f $ and compare them with the experimental data (Figure \ref{fig:SAPA}). The profiles manifest improvements in estimating the sediment erosion by the multi-resolution model (with $ q=4 $), while the single-resolution model underestimates the granular flow evolution. The remaining incompatibility between the numerical and experimental profiles (especially for the eroded area of case A-PVC at $ T\gtrsim3 $) originate from complex water-sediment mixing processes at the interface that the continuum-based particle method is incapable of simulating. Table \ref{tab:3} represents the normalized root-mean-square-error, $ L_2 (\cdot) $ of the sediment erosion parameters. The quantified results show a significant reduction in the numerical errors by the multi-resolution simulations (by a factor of $ \sim2-4 $) with respect to the single-resolution results. Moreover, to compare the global mechanical behavior of the sediment flow, we plot the global kinetic energy of the mixture particles ($ E_k=0.5\sum {\rho_{0}}_i{{l_0}_i}^2{\|\boldsymbol{v}\|_i}^2 $ for $ \forall i\in \Omega_m $) which is independent of the conditions used for detecting the eroded area (Figure \ref{fig:Ek}). For case A, $ E_k $ of the multi-resolution model (where $ q=4 $) is two to three times greater than $ E_k $ of the single-resolution model (where $ q=1 $) for both bed materials. Overall, we observe that the developed multi-resolution model considerably reduces the errors of the single-resolution simulations by estimating more sediment erosion at the interface. 

\begin{table}
	\begin{center}
		\caption{The normalized root-mean-square-error ($ L_2 $) of $ A_e $, $ {x_c}{A_e} $, and $ x_f $ with $ q=1 $, 2, and 4 for case A with sand and PVC.}
		\small
		\begin{tabular}{c c c c c c c c c c } 
			\hline
			Case A & \multicolumn{3}{c}{$ L_2(A_e) $\%} & \multicolumn{3}{c}{$L_2({x_c}{A_e})$\%}& \multicolumn{3}{c}{$L_2({x_f})$\%}\\
			\hline
			$ q $ & \multicolumn{1}{c}{$1$} &\multicolumn{1}{c}{$2 $}&\multicolumn{1}{c}{$ 4 $} &\multicolumn{1}{c}{$1$} &\multicolumn{1}{c}{$ 2 $}&\multicolumn{1}{c}{$4 $} &\multicolumn{1}{c}{$1$} &\multicolumn{1}{c}{$ 2 $}&\multicolumn{1}{c}{$4 $} \\
			\hline
			Sand & 32.01 & 20.10 & 8.49& 25.52&13.47&8.20& 9.88&4.33&2.40\\ \hline
			PVC &  35.41 & 30.64 & 13.77& 25.06&14.76&2.23& 5.08&1.29&1.39\\\hline
		\end{tabular}
	 
		\label{tab:3}
	\end{center}
\end{table}

\section{Conclusion}\label{sec:Concl_A5}
We investigated the mechanical behavior of fluid-driven granular dynamics through a consistent multi-resolution particle method. We developed a conservative form of governing equations (based on the enhanced weakly-compressible MPS method) to incorporate multi-scale water-sediment interactions within the continuum-based numerical modeling. Simulating two benchmark cases (i.e., the multi-viscosity Poiseuille flow and the multi-density hydrostatic pressure problems), we studied the accuracy and convergence of the numerical results with the single- and multi-resolution models. As for the fluid-driven granular erosion, we adopted the generalized rheology equation to model the two-dimensional dam-break waves on erodible sediment beds. We presented and discussed the mechanical behavior of this benchmark case for various configurations (i.e., with the flat-bed and the step-like discontinuities) and bed materials (i.e., the sand and PVC granules). We analyzed the sensitivity of sediment dynamics to the added suspension equation and the constant material parameters of the rheology model. Through comprehensive numerical validations, we studied the flow evolution and mechanical properties of the sediment erosion induced by the rapid water waves. Moreover, we compared the numerical results of the single- and multi-resolution simulations to evaluate the role of multi-scale interactions in capturing the global behavior of this benchmark case.

The particle convergence study on the two numerical benchmark cases confirms that the proposed multi-resolution formulation predicts the analytical results with acceptable accuracy. For the multi-viscosity Poiseuille flow, the multi-resolution shear force respects the convergence behavior of the numerical model and keeps the errors of the velocity profiles to less than 2 \%. For the multi-density hydrostatic pressure, the conservative governing equations ensure the accuracy and stability of the results with errors less than 0.5 \%. In both cases, increasing the spatial resolution, the numerical results converge to the analytical profiles with the convergence order greater than one. 

For the dam break waves on movable beds, the developed numerical model provides in-depth details of the water-sediment mixing processes and the global behavior of sediment dynamics. The nonlinear vertical velocity and granular concentration profiles of the flat-bed with PVC are well estimated. Simulating different geometrical setups with the sand and PVC granules clarifies that the sediment dynamic greatly depends on the mobility of the bed materials, i.e., the density ratio, $ \rho_m/\rho_w $ (supporting the theoretical analysis by \cite{Spinewine2013} and \cite{Fraccarollo2002}). In the flat-bed configuration, the eroded sediment and the thickness of the transport layer increase with the light PVC material (by a factor of $ \sim2.5 $) compared with the case with the sand grains. The step-like discontinuities lead to complex bed-load evolution close to the gate's location and slightly increase the mobility of the bed material at the initial stages of erosion. Regardless of the initial configuration, in cases with sand, the wave propagates on the horizontal bed with almost identical bed-load layer thickness and speed. In general, the multi-resolution MPS method proves to be capable of simulating the complex sediment erosion at the interface and the wavefront position. Despite the local discrepancies, the interface data and the flow evolution match well with the experimental measurements. Moreover, we observe that the multi-resolution model captures more interface deformations in comparison with the single-resolution model that underestimates sediment erosion. The results presented highlight the importance of multi-scale multiphase water-sediment interactions in numerical simulations of such a rapid fluid-driven problem. 

It is worthwhile to employ the developed particle method for studying complex multi-directional granular flows with applications to industrial and hydro-environmental problems. The proposed multi-resolution formulations can be extended to two-phase mixture models (e.g., \cite{Baumgarten&Kamrin2019} and \cite{Shi2019}) to investigate the effects of dilatation and compaction on the mechanical behavior of immersed granular flows. Furthermore, the multi-resolution MPS method can be further validated to high-density ratio problems and coupled with dynamic particle merging and splitting techniques to simulate violent free-surface flows \cite{Jandaghian2021_3,Rezavand2020} and fluid-structure interactions \cite{Zhang2020_FSI,Khayyer2019_FSI}. 
\section*{Supplementary data}{\label{SupMat} Movies of the simulations are available online.}
\section*{Acknowledgments}
Authors acknowledge the financial support of the Natural Sciences and Engineering Research Council of Canada (NSERC) and Polytechnique Montréal, Canada. This study used the high-performance computing resources of Compute Canada and Calcul Quebec.
\section*{Declaration of interests}{The authors report no conflict of interest.}
\section*{Author ORCID}
M. Jandaghian, https://orcid.org/0000-0001-5111-9640;

A. Shakibaeinia, https://orcid.org/0000-0001-8219-1469



 \bibliographystyle{elsarticle-num} 
\bibliography{References}





\end{document}